\documentclass[pr, twocolumn, preprintnumbers, showpacs,footnoteadded, superscriptaddress,nofootinbib,longbibliography]{revtex4-1}
\pdfoutput=1

\usepackage[T1]{fontenc}
\usepackage{amssymb}
\usepackage{graphicx}
\usepackage{amsmath}
\usepackage{hyperref}
\usepackage{tensor}
\usepackage{epstopdf}
\usepackage{multirow}
\usepackage{extarrows}
\usepackage{graphicx,subfigure}
\newcommand{\tabincell}[2]{
\begin{tabular}{@{}#1@{}}#2\end{tabular}
}

\newcommand{\td}{\mathrm{d}}
\newcommand{\te}{\mathrm{e}}

\begin{document}
\title{\boldmath Stability analysis on charged black hole with non-linear complex scalar}
\author{Zhan-Feng Mai}
\email{zhanfeng.mai@gmail.com}
\author{Run-Qiu Yang}
\email{aqiu@tju.edu.cn}
\affiliation{Center for Joint Quantum Studies and Department of Physics, School of Science, Tianjin University, Yaguan Road 135, Jinnan District, 300350 Tianjin, P.~R.~China}

%\pacs{PACS}
%\keywords{keywords}
%\preprint{preprint}
%%%%%%%%%%%%%%%%%%%%%%%%%%%%%%%%%%%%%%
%\begin{abstract}
%%%%%%%%%%%%%%%%%%%%%%%%%%%%%%%%%%%%%%
\begin{abstract}
It has been shown recently that the charged black hole can be scalarized if Maxwell field minimally couples with a complex scalar which has nonnegative nonlinear potential. We firstly prove that such scalarization cannot be a result of continuous phase transition for general scalar potential. Furthermore, we numerically find that it is possible that the RN black hole will be scalarized by a first order phase transition spontaneously and near extremal RN black hole is not stable in micro-canonical ensemble. In addition, considering a massless scalar perturbation, we compute the quasi-normal modes of the scalarized charged black hole and the results do not only imply that the spontaneously scalarized charged black hole is favored in thermodynamics but also suggest that it is kinetically stable against scalar perturbation at linear level. Our numerical results also definitely gives negative answer to Penrose-Gibbons conjecture and two new versions of Penrose inequality in charged case are suggested.
\end{abstract}

%\end{abstract}
%%%%%%%%%%%%%%%%%%%%%%%%%%%%%%%%%%%%%%
\maketitle
%\tableofcontents
\flushbottom

%%%%%%%%%%%%%%%%%%%%%%%%%%%%%%%%%%%%%%
\noindent

\section{Introduction}\label{intro1}
Recently, the observation of gravitational wave and black hole image provide a new motivation and vision in black hole physics\cite{Abbott:2016blz,Akiyama:2019cqa}. Providing a deep understanding in quantum gravity, black hole physics will always be a long-live project to study, theoretically and experimentally. One of the most important topics on black hole physics is no-hair theorem, which claims that black holes are determined only by mass, charge and angular momentum respectively\cite{Ruffini:1971bza}. It was first concluded by Bekenstein that the static and spherical neutral black hole in asymptotic flat spacetime cannot be endowed with real scalar field, Proca field and spin-2 field\cite{Bekenstein:1972ny}. Furthermore, the no-hair theorem was extended by Mayo and Bekestein that the static and spherical black hole cannot be endowed with a coupling charged scalar field together with a non-negative self-interacting potential \cite{Mayo:1996mv}. However, the requirements of above no-hair theorems are as strong as its conclusion. Therefore, it is not difficult to find hairy black hole solutions if one breaks the requirements of these no-hair theorems. In Refs.~\cite{Feng:2013tza,Liu:2013gja,Fan:2015tua,Khodadi:2020jij}, taking account into a non positive definite potential or asymptotic AdS spacetime, $D$-dimension scalarized black holes have been constructed in Einstein theory minimally coupled with neutral scalar field. Recently, in the extended Maxwell theory non-minimally coupled with a scalar field, such as Einstein-Maxwell-Dilaton (EMD) theory\cite{Fan:2015oca}, Einstein-Maxwell-Scalar theory\cite{Konoplya:2019goy}, Einstein-Born-Infeld (EBI) theory\cite{Wang:2020ohb,Stefanov:2007eq}, and Quasi-topological Electromagnetism theory\cite{Myung:2020ctt}, a class of charged black hole with scalar hair has been found. For a recent review relevant to the no-hair theorem, See Ref.~\cite{Herdeiro:2015waa}.

The above examples give the simplest case of the black hole with an additional scalar hair. As we have mentioned, the Bekenstein's no-scalar-hair theorem was generalized by Ref.~\cite{Mayo:1996mv}, claiming a very strong conclusion: the non-extremal static and spherical charged black hole cannot carry charged scalar hair whether minimally or nonminimally coupled to gravity, and with a regular positive semidefinite self-interaction potential. However, a numerical charged black hole solution with Q-hair was found in Einstein-Maxwell gravity minimally coupled with a non-linear complex scalar, the self interacting potential taking the following polynomial function \cite{Hong:2020miv}
\begin{equation}\label{defVs1}
V(|\psi|^2)=\frac{m^2}{2}|\psi|^2-\frac{\lambda}{4}|\psi|^4 + \frac{\beta}{6}|\psi|^6\,.
\end{equation}
It was also pointed out that the detailed form of the potential $V(|\psi|^2)$ is not crucial and the Q-hair can exist for large class of nonlinear potential\cite{Hong:2019mcj,Herdeiro:2020xmb}. The reason why Ref.~\cite{Mayo:1996mv} obtained a wrong statement is that it omitted a scalar mass term at an asymptotic infinity. Though the black holes with scalar hair has been found in many physical models, the discovery of Refs.~\cite{Hong:2020miv} has a few of special interesting aspects. Particularly, in this model, the gravity, Maxwell field and complex scalar field are all minimally coupled with each other and the potential of scalar field is positive semidefinite. This gives a possibility to realize the scalar hairy black hole in the Einstein gravity and asymptotically flat spacetime. Furthermore, as the potential of scalar field is nonnegative, such model will have stable true vacuum state $\psi=0$ in Minkowski spacetime.

It needs to note that the Reissner-Nordstr\"{o}m black hole (RN black hole) is still a solution of field equations even if the Q-hair appears in the models discussed by Refs.~\cite{Hong:2020miv,Hong:2019mcj}. Given a scalarized charged black hole solution, it is worth to investigate whether the scalarized charged black hole is more stable than RN black hole in thermodynamics, i.e. whether the RN black hole can be spontaneously scalarized by phase transition. Recently, it has been well studied the thermodynamic self-stability associated with heat capacity in various ensembles in \cite{Caldarelli:1999xj,Mo:2013sxa,Zhang:2018rlv,Quevedo:2006xk,Quevedo:2013pba}. However, we focus on  investigating the thermodynamic stability of scalarized charged black holes, compared with the RN black hole in various ensembles. Concretely, in microcanonical ensemble, given the same ADM mass $M$ and total charge $Q$, the stability requires the maximum of the black hole entropy $S(M,Q)$ and a phase is more stable than the other if it has larger entropy. In canonical ensemble, with fixing Hawking temperature $T$ and $Q$, the black hole owning less Helmholtz free energy $F(T,Q)$ will be the more stable, while in grand canonical ensemble, with the identical $T$ and chemical potentials $\mu$, the black hole which has smaller Gibbs free energy $G(T,\mu)$ indicates it will be more stable. Furthermore, the black hole entropy $S(M,Q)$ is given by the area of horizon according to Bekenstein's entropy formula while the thermodynamic potentials $F(T,Q)$ and $G(T, \mu)$ can be read off from partition function via Euclidean path-integral approach developed by Hawking and York, et al. \cite{Gibbons:1976ue,Brown:1989fa,Braden:1990hw}.

Remarkably, in astrophysics, a real black hole practically is more closer to grand canonical ensemble allowing charge and energy exchange with other matter in universe. However, as theoretical research interest, we still take the case of microcanonical ensemble and canonical ensemble under our consideration to study whether the RN black hole will be scalarzied under the process of discontinuous phase transition spontaneously.
Furthermore, by means of analyzing the corresponding entropy or thermodynamic potentials in various ensembles, we find that there is a possibility that the scalarized charged black hole is more stable than the RN black hole in thermodynamics. This implies that, for a large class of nonlinear complex model, it is possible that the RN black hole will be scalarized spontaneously via a first order phase transition. Our numerical results also imply that the scalarized black hole is always more stable than the RN black hole in microcanonical ensemble and when $M\rightarrow|Q|$. This implies the isolated near extremal RN black hole is not stable and will spontaneously scalarize.

In addition to the thermodynamic stability, another natural question is whether the scalarized charge black hole is kinetically stable, i.e. stable under against a small perturbation at least at linear level. It has been developed by Vishveshwara and Teukolsky et al.\cite{Teukolsky:1973ha,Press:1973zz,Vishveshwara:1970cc} that in static axial or spherical symmetric background, the equation of motion associated with the perturbation field can reduce to radial equation in frequency domain. Furthermore, the radial equation can be interpreted as an eigenvalue problem. Specifically, imposing a physical boundary condition both in the spacial infinity and at the black hole horizon, a class of complex frequency called black hole quasi-normal modes(QNMs), which implies dissipation at both event horizon and spatial infinity, can be picked out. The instability of black hole might be triggered due to the negative imaginary part of the complex frequency. In general, the QNMs can be solved by shooting method in numerics. Approximatively, there have been some analytical approaches to achieve to calculate QNMs as well: the WKB approximation method, continue fraction method and Monodromy method. However, recently it has been argued in Ref.~\cite{Konoplya:2019hlu} that the WKB approximation method cannot catch the unstable mode within the spectrum of the QNMs. Furthermore, we also find that the shooting method cannot efficiently give the stable modes through numerical error analysis. We thus adopt hybrid method to calculate the QNMs for studying the kinetic stability of scalarized charged black hole. If a given solution is not kinetically stable, then such solution cannot exist in a real physical system. For the situations that the scalarized black hole is more thermodynamically stable than RN black hole, but it does not automatically guarantee that the scalarized charged black hole is still stable kinetically. However, our results show the neutral perturbative scalar field will not trigger the instability of the scalarized charged black hole at linear level. In addition, for a recent review on perturbation theory on black hole, stability analysis and quasi-normal modes of black hole, more detail has been given in Refs.~\cite{Pani:2013pma, Konoplya:2011qq, Berti:2009kk}.

This paper will be organized as follow: In Sec.~\ref{modelset}, we briefly introduce the model of Einstein-Maxwell theory minimally coupled with a non-linear complex field. For general non-linear semi-definite potential, we present a proof that the scalarized charged black hole cannot be a result of continuously scalarization. In Sec.~\ref{thermo}, Giving a logarithmic potential, we obtain a class of numerical scalarized charged black hole solution and in various ensembles, investigate their thermodynamic stability compared with the RN black hole. We find that in both microcanonical ensemble and canonical ensemble, it is possible that the RN black hole can be spontaneously scalarized via first order phase transition. In Sec.~\ref{KietSta}, considering a probing neutral scalar field, we investigate the stability on the scalarized charged black hole by means of computing the quasi-normal modes using both the shooting method and the WKB approximation method. This suggests that the probing neutral scalar field cannot trigger the instability of the sclarized charged black hole. In Sec.~\ref{Conclu}, we present our conclusion and further discussions.

\section{Model Setup}\label{modelset}
In this paper, we consider the following Einstein-Maxwell theory minimally coupled with a non-linear complex scalar field (we set $G_N=c_s=\hbar=k=1$, $c_s$ denotes the speed of light),
\begin{eqnarray}\label{action1}
&&S=\frac1{16\pi}\int\td^4x\sqrt{-g} {\cal L}  \, , \cr
~\cr
&&{\cal L}=\left(R-F_{\mu\nu}F^{\mu\nu}-(D_\mu\psi)^\dagger(D^\mu\psi)-W(|\psi|^2)\right) \, ,
\end{eqnarray}
where the covariant derivative operator $D_{\mu}=\nabla_\mu-iqA_\mu$, $q$ denotes charge of the complex scalar field.  In addition, $F_{\mu\nu}=\partial_\mu A_{\nu}-\partial_\nu A_{\mu}$ is the strength tensor of the U(1) electromagnetic field $A_{\mu}$. The non-linear positively semi-definite potential $W(x)$ is a smooth function and satisfies following requirements:
\begin{equation}\label{reqw1}
W(0)=0,~~W'(0)=m^2>0,~~W(x)>0~\text{if}~x>0\,.
\end{equation}
These conditions insure that scalar field has stable true vacuum $\psi=0$ in Minkowski spacetime when $F_{\mu\nu}=0$. Performing variation on the action \eqref{action1} with respect to the metric $g_{\mu\nu}$, the gauge field $A_\mu$ and the complex scalar field $\psi$ respectively, we obtain the equation of motion
\begin{eqnarray}\label{eom}
&&R_{\mu\nu}-\frac{R}2g_{\mu\nu}=8\pi [T^{(M)}_{\mu\nu}+T^{(\psi)}_{\mu\nu}]\,,  \cr
&& ~ \cr
&&\nabla^\mu F_{\mu\nu}=iq\left(\psi^\dagger D_\mu\psi-\psi (D_\mu\psi)^\dagger\right) \, , \cr
&& ~ \cr
&&D^2\psi-w(|\psi|^2)\psi=0\, ,
\end{eqnarray}
where we define $w(x)=W'(x)$ and the energy momentum tensor associated with both the electric field and the complex scalar field reads
\begin{eqnarray}\label{Energy}
&&T^{(M)}_{\mu\nu}=\frac1{4\pi}\left(F_{\mu\sigma}{F^{\sigma}}_\nu-\frac14g_{\mu\nu}F_{\rho\sigma}F^{\rho\sigma}\right)\, , \cr
~\cr
&&T^{(\psi)}_{\mu\nu}=\frac1{16\pi}
\left( D_\mu\psi(D_\nu\psi)^\dagger+D_\nu\psi(D_\mu\psi)^\dagger \right. \cr
&&\left.-g_{\mu\nu}\left((D_\mu\psi)^\dagger(D^\mu\psi)+W(|\psi|^2)\right)\right )\, .
 \end{eqnarray}
For the non-linear potential, the requirement~\eqref{reqw1} implies that
\begin{equation}\label{smallxw}
w(x)=m^2+w_1x+w_2x^2+\cdots
\end{equation}
with some constants $\{w_1,w_2,\cdots\}$  when $|x|\ll1$.

To obtain the scalary charged black hole and investigate some related properties, we adopt the following spherical symmetric line element,
\begin{equation}\label{metric0}
  \td s^2=-f(r)\te^{-\chi(r)}\td t^2+\frac{\td r^2}{f(r)}+r^2(\td \theta^2 + \sin^2 \theta \td \varphi^2).
\end{equation}
Due to the spherical symmetry, the electromagnetic field and the complex scalar field take the following form,
\begin{equation}\label{matters1}
  A_\mu=\phi(r)(\td t)_\mu,~~\psi=\psi(r)\,.
\end{equation}
From the line element Eq.~\eqref{metric0}, the Hawking temperature $T$ reads
\begin{equation}\label{hawkT1}
  T=\frac{f'(r_h)\te^{-\chi(r_h)/2}}{4\pi}\, ,
\end{equation}
where $r_h$ denotes the outer event horizon.
With the given spherical anstaz, the equations of motion reduce to
\begin{equation}\label{eqscalar1}
\begin{split}
  &\psi''+\left(\frac{f'}{f}-\frac{\chi'}2+\frac2r\right)\psi'+\left(\frac{\te^{\chi}\phi^2q^2}{f^2}-\frac{w(\psi^2)}{f}\right)\psi=0\\
  &\phi''+(\chi'/2+2/r)\phi'-\frac{\psi^2q^2}{2f}\phi=0\,,\\
  &\chi'+\frac{r\te^{\chi}\psi^2\phi^2q^2}{f^2}+r\psi'^2=0\,,\\
  &f'+\left(\frac1r+\frac{r\psi'^2}2\right)f+r\te^{\chi}\phi'^2 \\
  &-\frac1r+\frac{\te^{\chi}r\psi^2\phi^2q^2}{2f}+\frac12rW(|\psi|^2)=0\, ,
  \end{split}
\end{equation}
where the prime denotes the derivative with respect to $r$. In this paper, we consider the asymptotically flat space time. The scalar field, gauge field and metric components should satisfy the following regular boundary conditions when $r\rightarrow\infty$
\begin{eqnarray}\label{bdcond1}
  &f=1-\frac{2M}r+\cdots,~~\chi=\frac{\chi_2}{r^2}+\cdots \, , \cr
  ~\cr
  &\phi=\mu-\frac{Q}{r}+\cdots,~~|\psi|\leq\mathcal{O}(1/r^2)\, ,
\end{eqnarray}
where $\mu$ is the chemical potential, $M$ is the ADM mass and $Q$ is the total charge. With this boundary conditions and noting the fact that $w(|\psi|^2)\rightarrow m^2$ near the boundary, we find that the first equation of Eq.~\eqref{eqscalar1} reduces into following simple form near the infinity
\begin{equation}\label{eqscalar2}
  \psi''+\frac{2\psi'}r+(q^2\mu^2-m^2)\psi=0\, .
\end{equation}
The solution reads
\begin{equation}\label{asysol1}
  \psi(r)=\frac{\psi_+}{r}\te^{r\sqrt{m^2-q^2\mu^2}}+\frac{\psi_-}r\te^{-r\sqrt{m^2-q^2\mu^2}}\,.
\end{equation}
The boundary conditions Eq.~\eqref{bdcond1} implies following constraints
\begin{equation}\label{constraint1}
  m^2-q^2\mu^2>0,~~\psi_+=0\,.
\end{equation}
These are two necessary conditions on the spontaneous scalarization for asymptotically flat black holes.

Taking into account a polynomial potential, as mentioned above, a numerical charged black hole solution with Q-hair has been found in \cite{Mayo:1996mv, Hong:2020miv}. A natural question arises whether this class of scalarized black hole solution will arise from continuously spontaneous scalarization for specific non-linear potential. This question is important because if the answer is yes, the spacetime geometry can transit into scalarized black hole smoothly, otherwise, the latent heat will be relaxed or absorbed during phase transition between the RN black hole and scalarized black hole. Moreover, such latent heat will leave some observable effects if such a phase transition happened in our universe.

\begin{figure}[htpb]
  \centering
  \includegraphics[width=0.35\textwidth]{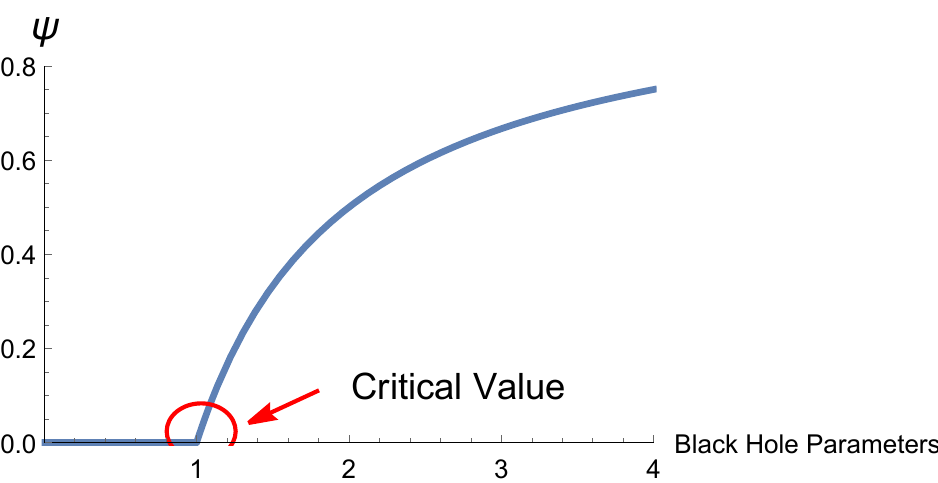}
  \includegraphics[width=0.35\textwidth]{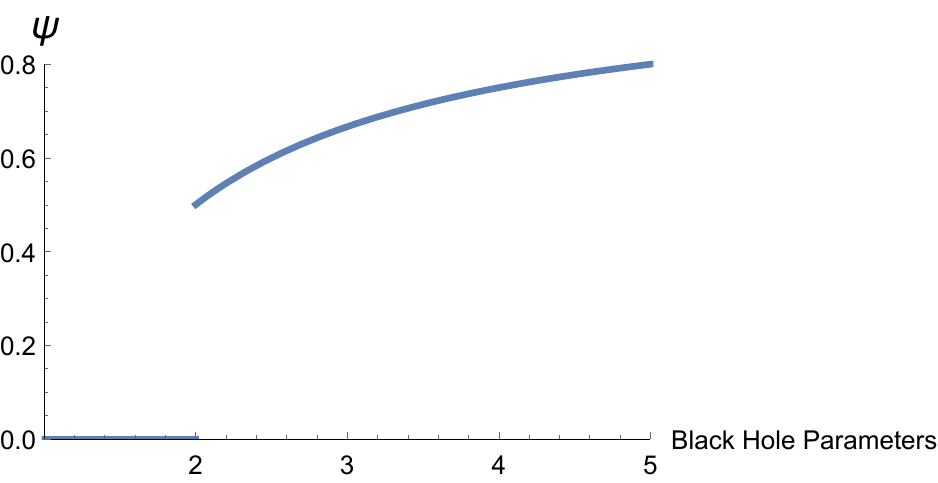}
 \caption{\textbf{Top}: Scalarization appear via continuous phase transition. \textbf{Bottom}: Scalarization appear via discontinuous phase transition.}\label{conscal}
\end{figure}

In the following, we shall present a proof that for arbitrary non-linear potential $W(|\psi|^2)$ satisfying the requirement~\eqref{reqw1}, the spontaneous scalarization of RN black hole cannot happen via continuous phase transition. As shown in Fig.~\ref{conscal}, if such continuous phase transition can happen, when we tune the parameters (total charge, chemical potential, temperature, etc.) of the black hole, there is a critical point where the strength of scalar field begins to increase into nonzero continuously. In other words, there must be a small region near the critical value associated with the black hole parameters (See also the top panel of Fig.~\ref{conscal}) where the complex scalar field is infinitesimal.
%
%\begin{theorem}\label{theorem1}
%The scalar hair cannot increase continuously from zero to nonzero if non-linear potential $W(|\psi|^2)$ satisfies requirement~\eqref{reqw1}.
%\end{theorem}
%\textit{Proof}:
Without losing generality, we can assume this infinitesimal scalar field has following form,
\begin{equation}
\psi(r, \theta, \varphi)=\varepsilon \sum_{l,\nu} a_{l \nu} (r) Y_{l \nu} (\theta, \varphi)\, \quad \varepsilon\rightarrow0\, .
\end{equation}
where $Y_{l \nu}$ denotes the spherical harmonic function, $\nu=0, \pm 1, \pm 2, \cdots$ is the magnetic quantum number and $l > \nu$ is the azimuthal quantum number.
Taking it into Eq.~\eqref{eom} and neglecting all the non-linear terms of $\varepsilon$, we find that the  metric and gauge field  decouple from the scalar field. Then spacetime geometry and gauge field are given by a RN solution
\begin{eqnarray}\label{rnsolu1}
  &\chi=0,~~f(r)=\frac{(r-r_h)(r-\mu^2r_h)}{r^2} \, , \cr
  ~\cr
  &\phi=\mu(1-r_h/r),~~\mu\in[-1,1]\,.
\end{eqnarray}
Here we require $\mu\in[-1,1]$ since the $r_h$ is defined to be the most outer horizon. The E.O.M of scalar field reads
\begin{equation}\label{eqscalar2}
  a_{l\nu}''+\left(\frac{f'}{f}+\frac2r\right)a'+\left(\frac{\phi^2q^2}{f^2}-\frac{m^2}{f}-\frac{l(l+1)}{f r^2}\right)a_{l\nu}=0 \, .
\end{equation}
This can be rewritten into following form
\begin{equation}\label{eqscalar3}
  \frac{\td}{\td r}\left(r^2f a_{l \nu}\frac{\td a_{l \nu}}{\td r}\right)=\left[\left(\frac{\tilde{m}(r)^2}{f}-\frac{\phi^2q^2}{f^2}\right)a_{l \nu}^2+a_{l \nu}'^2\right]r^2f \, .
\end{equation}
where we have redefine the effective mass as $\tilde{m}(r)^2 = m(r)^2 + l(l+1)/r^2$ and it is obviously that $\tilde{m}^2 \geq m^2$.
Integrate it from horizon to infinity and we find
\begin{equation}\label{eqscalar4}
  \left.r^2f a_{l \nu}\frac{\td a_{l \nu}}{\td r}\right|_{r_h}^\infty=\int_{r_h}^{\infty}\left[\left(\frac{\tilde{m}(r)^2}{f}-\frac{\phi^2q^2}{f^2}\right)a_{l \nu}^2+a_{l \nu}'^2\right]r^2f\td r \, .
\end{equation}
As $a(r)$ is a regular at horizon and decays to zero at infinity, the left side of Eq.~\eqref{eqscalar4} is zero. Thus, we have
\begin{equation}
\int_{r_h}^{\infty}\left[\left(\frac{\tilde{m}(r)^2}{f}-\frac{\phi^2q^2}{f^2}\right)a_{l \nu}^2+a_{l \nu}'^2\right]r^2f\td r=0\,.
\end{equation}
This means that $\exists r_0\in(r_h,\infty)$ such that
\begin{equation}\label{condh1}
  \frac{\tilde{m}(r)^2}{f(r_0)}-\frac{\phi(r_0)^2 q^2}{f(r_0)^2}<0\,.
\end{equation}
We then obtain
\begin{equation}\label{condh2}
m^2\leq \tilde{m}(r)^2<\frac{\phi(r_0)^2 q^2}{f(r_0)}=\mu^2q^2\frac{r_0-r_h}{r_0-\mu^2r_h}<\mu^2q^2\,.
\end{equation}
Here we have used the solution~\eqref{rnsolu1}. Above result and requirement~\eqref{constraint1} are contradictory. This shows that, if the complex scalar field appears in a static black hole, its strength cannot be infinitesimal no matter how we choose the black hole parameters.  Thus, the continuous phase transition from RN black hole to scalarized charged black cannot occur.
%$\square$.

The demonstration above implies that the scalarized charged black hole solution cannot produce from the continuous spontaneous phase transition. However, the spontaneous scalarization on charged black hole may arise through non-continuous phase transition, namely the first-order phase transition. In this case, the complex scalar field jumps into nonzero from zero when we tune the parameters of black hole, see the bottom panel of Fig.~\ref{conscal}. Recall the equivalence between the black hole system and the thermal mechanic system, it is thus intriguing to study the thermodynamic stability of the scalarized charged black hole, compared with the corresponding charged black hole with no hair. Motivated by this, interpreted charged black hole with scalar hair as a class of ensemble, in the following, we focus on investigating the thermodynamical stability on scalarized charged black hole in various ensembles: microcanonical ensemble, canonical ensemble and grand canonical ensemble, and try to check if there is a first order phase transition.

\section{Thermodynamic Instability}\label{thermo}
In this section, we analyze the thermodynamic stability of scalarized charged black hole by proposing a specific example. As mentioned, the relation of Beskein entropy implies that black holes can be investigated as a thermodynamical system in term of three different ensemble: microcanonical ensemble, canonical ensemble and grand canonical ensemble. The microcanonical ensemble implies that black holes are interpreted as an isolated system where there does not exist any charge and energy exchange of black holes. The canonical ensemble indicates that black holes only exchange energy with environment of which the temperature is fixed. The grand canonical ensemble is similar to canonical ensemble but also admits the black hole to exchange the particles with environment. In practice, black holes in astrophysics are more likely to be grand canonical ensemble with exchange of particles and energy. However, as theoretical investigation on scalaized charged black hole, in following we still take microcanonical ensemble, canonical ensemble and grand canonical ensemble into account, studying the thermodynamical stability associated with the scalarized charged black hole and RN black hole respectively. Furthermore, we find that in grand canonical ensemble, the RN black hole is more stable than the scalarized charged black hole in thermodynamics, corresponding to general expectation. Nevertheless, we also find that the discontinuous scarization on RN black hole may happen in both microcanonical ensemble and canonical ensemble respectively. In the following, we will give more detail discussion.

Before go on discussing our result, let us first make short comment on different ensembles in black holes. The Euclidean path-integral approach to black hole thermodynamics originally proposed by Hawking in microcanonical ensemble~\cite{Hawking:1976de}. Later on, the canonical ensemble was investigated by York et al.~\cite{York:1986it,Whiting:1988qr,Brown:1994su}. It was found that suitable boundary conditions must be added in canonical ensemble. Then the York's approach was generalized into other ensembles such as the charged black hole in the grand canonical ensemble~\cite{PhysRevD.42.3376,Brown_1990}. It has been found that the results obtained by using the path-integral approach depend on the boundary conditions~\cite{Brown:1994gs,Hawking:1982dh}. The stability of black holes then also depends on the choice of boundary conditions and, consequently, on the choice of ensembles~\cite{Comer_1992}. In fact, the stability properties of a black hole are drastically influenced by the boundary conditions that determine ensemble.

In following we proceed to our discussion. We specify the non-linear potential as a logarithmic function with respect to the scalar field
\begin{equation} \label{potential}
W(x):=m^2 c^2 \log(1+\frac{x^2}{c^2}),
\end{equation}
where $c$ is a constant and $m$ is the effective mass of the scalar field. Considering the flat directions in gauge-mediated supersymmetric model~\cite{deGouvea:1997afu, Kusenko:1997si}, this potential is proposed in Ref.~\cite{Hong:2019mcj}, in which the supersymmetric breaking has been absorbed into the rescaling of $\psi$. In addition to satisfying the stable vacuum requirement Eq.(\ref{reqw1}) , the shape of potential is asymptotic flat when taking large field limit $\psi \gg c$. From the equation of motion of $\psi$ in Eq.~\eqref{eqscalar1}, one will find that the scalar field becomes massless large field limit $\psi \gg c$. Moreover, the logarithmic potential can bring better numerical stability as well.

Since this paper tries to find the black hole solutions with scalar hair, there should be a horizon locating at position $r=r_h$. In static spherically symmetric case, this implies $f(r_h)=0$. To set up the numerical method to solve the equation of motion Eq.~(\ref{eqscalar1}), we in principle still needs five additional independent boundary conditions at horizon. Practically, the regularity of physical fields at $r=r_h$ implies that the solution can approximatively be written as the following Taylor's series with respect to $r-r_h$,
\begin{eqnarray} \label{horbou1}
&&f(r)=f_1 (r-r_h)+\cdots, \quad \chi(r)=\chi_0+\chi_1(r-r_h)+\cdots \, ,\cr
~\cr
&&\phi(r)=\phi_0+\phi_1(r-r_h)+\phi_2 (r-r_h)^2+\cdots\, , \cr
~\cr
&&\psi(r)=\psi_0+\psi_1(r-r_h)+\cdots \, .
\end{eqnarray}
Take them into Eq.(\ref{eqscalar1}) and we will find
\begin{eqnarray} \label{horbou2}
&&\psi_1=\frac{2\psi_0 c^2 m^2}{(\psi_0^2+c^2)f_1}\, ,\quad \phi_0=0\,.
\end{eqnarray}
This leaves three independent variables $\{\psi_0, \chi_0, \phi_1\}$ at horizon and so we have 7 independent parameters in solving Eq.~\eqref{eqscalar1}
\begin{equation}
\{r_h, \psi_0, \chi_0, \phi_1, c, m, q \} \, ,
\end{equation}
in which $\{r_h, \psi_0, \chi_0, \phi_1 \}$ come from the value of various fields at the horizon, $q$ is the charge of the scalar field and $\{c, m\}$ are the parameters of the non-linear potential. It is remarkable that there is a scaling symmetry of the equation of motion
\begin{equation}
r \to \lambda r \, , \quad t \to \lambda t \, , \quad  q \to \frac{q}{\lambda} \, , \quad m \to \frac{m}{\lambda},
\end{equation}
which equivalently rescale the metric $g_{\mu\nu} \to \lambda ^2 g_{\mu\nu}$ and the electronic field $A_\mu \to \lambda A_\mu$. Due to such a symmetry, we fix $m=0.01$ for convenience. In addition, we must impose
\begin{equation}
\chi \to 0, \quad \text{when} \quad r \to \infty,
\end{equation}
which is a requirement about the normalisation of $t$ associated with the gravitational redshift\cite{Hartnoll:2008kx}. Practically, given any arbitrary value of $\chi_0$, the equation of motion is invariant when performing the following scaling transformation
\begin{equation} \label{tresc}
t \to \te^{-\frac{\chi_{\infty}}{2}}t \, , \quad \chi  \to \chi - \chi_{\infty} \, , \quad \phi(r) \to  \te^{\frac{\chi_{\infty}}{2}}\phi(r),
\end{equation}
where $\chi_{\infty}$ denotes the value of the solution $\chi(r)$ in the asymptotic infinity. In other words, we perform a time rescaling $t \to e^{-\frac{\chi_{\infty}}{2}}t$ to set $\chi=0$ at the boundary. In order to numerically solve the equations simply, we in general set $\chi_0=0$. However, such choice will lead to $\chi(\infty)=\chi_\infty\neq0$, we can thus finally transform the solution to satisfy $\chi(\infty)=0$ by the transformation~\eqref{tresc}.

Base on these two symmetries, two physical parameters, the charge of the complex scalar field $q$ and the parameter related to the scalar potential $c$, and three parameters as the initial value at the horizon $r_h$, $\phi_1$ and $\psi_0$ are left. Therefore, The integration of the equation of motion Eq.~\eqref{eqscalar1} from the event horizon to the infinity will give us a map:
\begin{equation} \label{map1}
\{r_h, \phi_1, \psi_0 , c, q \} \mapsto \{\mu, T, Q, M, \psi_+ \}\, .
\end{equation}
If one chooses five parameters $\{r_h, \phi_1, \psi_0 , c, q \}$ arbitrarily, the $\psi_+$ may be or may not be vanish. From Eq.~\eqref{constraint1}, we know that only the one satisfying $\psi_+=0$ is an admitted solution, leading to only four of parameters $\{r_h, \phi_1, \psi_0 , c, q \}$ are independent.

\begin{figure}[htpb]
  \centering
  \includegraphics[width=0.22\textwidth]{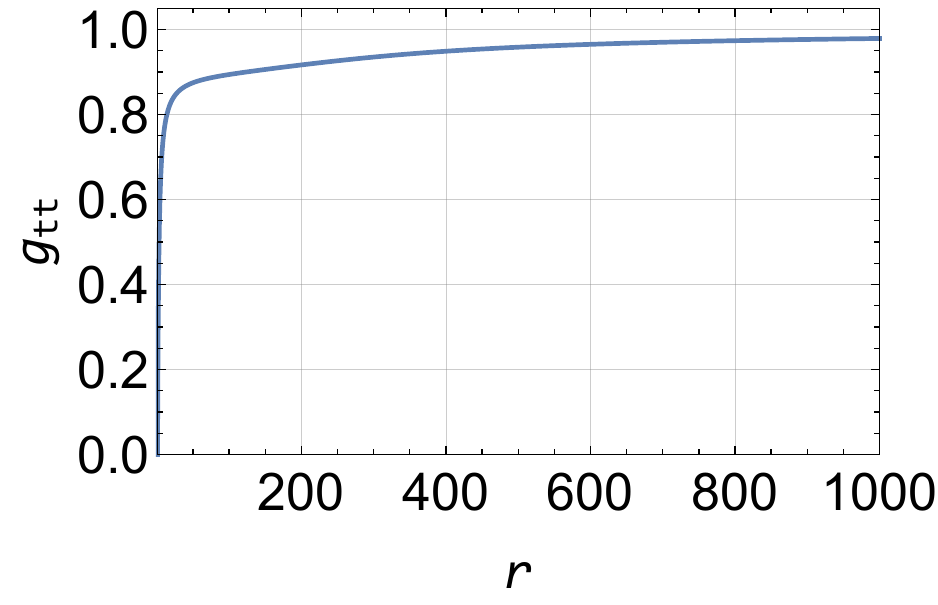}
  \includegraphics[width=0.22\textwidth]{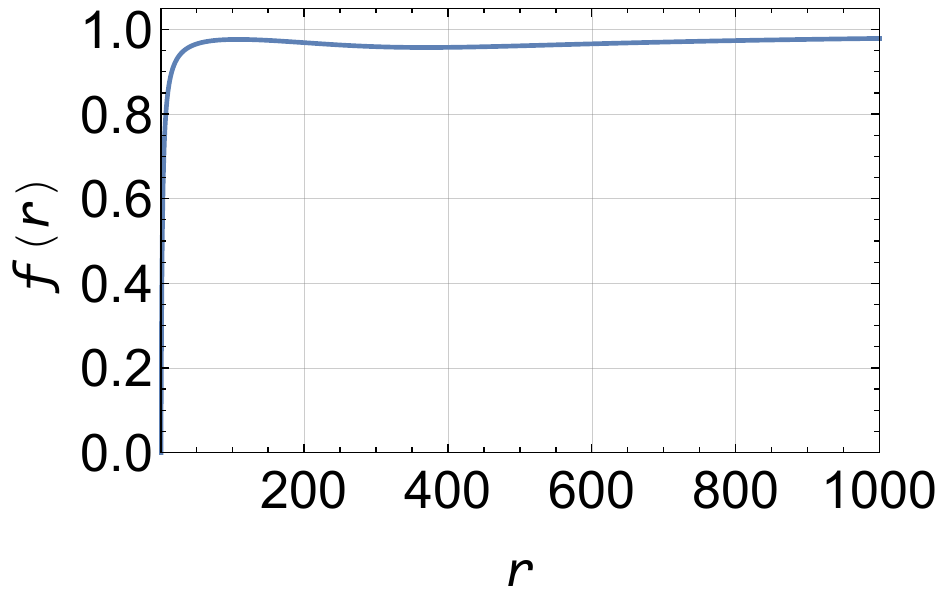}
  \includegraphics[width=0.22\textwidth]{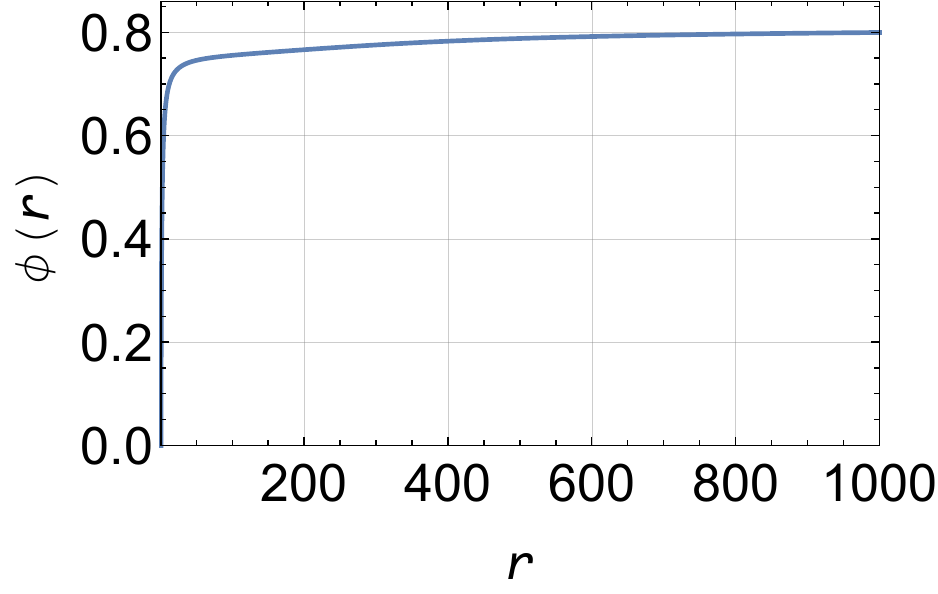}
  \includegraphics[width=0.22\textwidth]{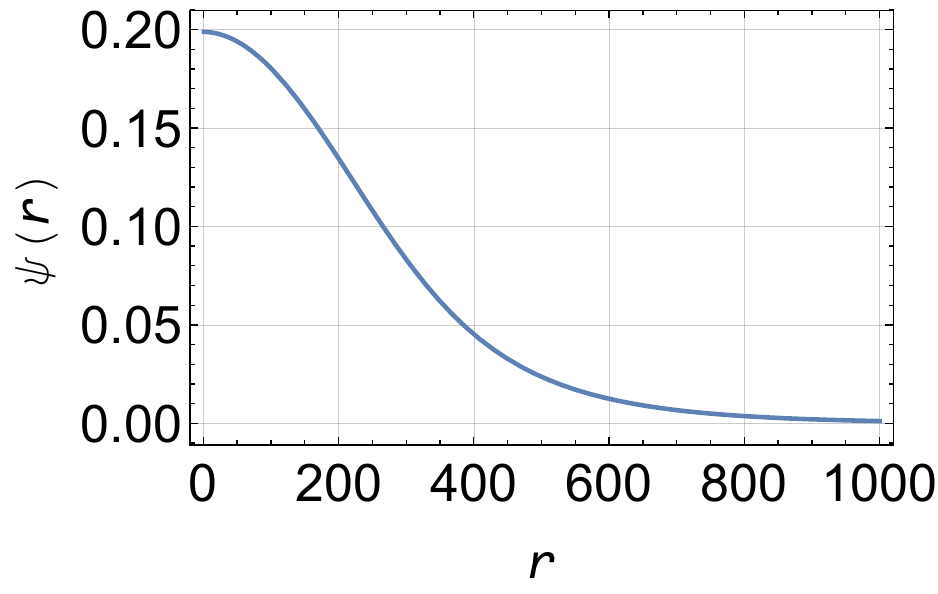}
 \caption{Numerical solutions of the metric components $g_{tt}=f(r)\te^{-\chi(r)}$, $1/g_{rr}=f(r)$, the electric potential $\phi(r)$ and the scalar field $\psi (r)$ respectively. Here we take parameters \{$r_h=1, c=0.1, \phi_1=0.8, m=0.01, q=\frac{11}{10}m$\}. The $\psi_0\approx0.1988$ is given by shooting method. }\label{fig1}
\end{figure}

We choosing the parameters as $r_h=1, c=0.1, \phi_1=0.8, m=0.01, q=\frac{11}{10}m$ and set $r=r_{\infty}=1000$ as the infinity cut off, finding that the complex scalar $\psi$ has been efficiently decay to $0$ at the infinity cutoff $r = 1000$, we firstly numerically solve the equation of motion Eq.~\eqref{eqscalar1} by Runge-Kutta methods under the initial condition Eq.~\eqref{horbou1} and Eq.~\eqref{horbou2}. As we have explained, the $\psi_0$ now is not free because we need to satisfy $\psi_+=0$. Then we use the standard shooting method to find the smallest value of $|\psi_0|$ to satisfy the constrain Eq.~\eqref{constraint1}. We find a numerical charged black hole solution with scalar hair when $\psi_0\approx0.1988$. After Performing the scaling transformation Eq.~\eqref{tresc}, we show the numerical solution are shown in Fig.~\ref{fig1} from which we find that $g_{tt}=f(r)\te^{-\chi(r)}$ is still a monotone increasing function with respect with $r$, indicating that in scalarized charged black hole, the gravity is still attraction, as the case of the RN black hole.

In addition, taking $r =1000$ as an infinity cutoff, from Fig.~\ref{fig1} one can see that the scalar field $\psi(r)$ have efficiently decay to $0$ smoothly at $r=1000$, implying that the spacetime manifold has efficiently reduce to RN black hole. In addition, if increase the infinity cutoff, to maintain effective numerical accuracy, one must increase the working precise, leading to increase the computational time. In practice, the best cut-off is chosen by following way: in double float accuracy, we take $\max r=2^6,2^7,2^8,2^9$ and $2^{10}\approx1000$. We observed that the differences of $\psi_0$ are smaller and smaller. However, if we increase the cut-off to be $2^{11}$ and more, we found the differences of $\psi_0$ are larger and larger. This implies 1000 is best cut-off. If we using quadruple float accuracy, the best cut-off is around 2000, however, the computational time will increase more 10 times. Therefore, for investigating the stability of scalarized charged black hole effectively and efficiently, it is reasonable to set $r=1000$ as an infinity cutoff.
%scalar-f1.pdf
%scalar1-chi1.pdf
%scalar-a1.pdf
%scalar1.pdf
%%%%%%%%%%%%%%%%%%%%%%%%%%%%%%%%%%%%%%%%%%%%%%%%%%%%%%%%%%%%%%%%%%%%%%%%%%%%%%%%%%%%%%%%%%%%%%%%%%%%%5

\subsection{Microcanonical Ensemble} \label{micensem}
\subsubsection{Thermodynamical Stability for scalarized charged black hole}
To compare which one is more stable, RN black hole or scalarized black hole, we need to specify what ensemble we will consider. We first consider the microcanonical ensemble describing an isolated system.  In the microcanonical ensemble, the characteristic thermodynamic variables is entropy and a physical realistic process will be always towards the direction of increasing entropy, a phase transition thus can happen only if the entropy will be increased. Therefore, in the case microcanonical ensemble, we consider an isolated black hole where the total mass $M$ and total charge $Q$ are fixed parameters and the black hole entropy can be interpreted as $S=S(M,Q)$. Since the black hole entropy is proportion to the area of black hole $S \sim 4\pi r_h^2$, which implies the larger the black hole radius indicates more stable in microcanonical ensemble, we need to compare horizon radii of scalarized black hole and RN black hole for the same mass $M$ and total charge $Q$.

Following our illustration above, we numerically analyze the behavior of the event horizon radius working with $(Q/M, \Delta r_h/M)$ plane with fixing total mass $M$, where $\Delta r_h=r_{h}-r_{h_{\text{RN}}}$ denotes difference between the scalarized charged black hole and the corresponding charged black hole with the same $M$ and $Q$.  Specifically, given a numerical scalarized charged black hole solution, the ADM mass and the total charge read
\begin{equation}\label{MQ}
M=-\frac{r}{2}(f(r)-1)|_{r \to r_{\infty}} \, , \quad Q= r^2 \phi (r) |_{r \to r_{\infty}}.
\end{equation}
Then, the event horizon radius of the corresponding RN black hole is
\begin{equation}
r_{h_{\text{RN}}}=M+\sqrt{M^2-Q^2}.
\end{equation}
Naively, we firstly consider the solution given in Fig.~\ref{fig1}. The mass and the total charge directly read $M \approx 11.53$ and $Q \approx 11.81$, implying that there does not exist the corresponding RN black hole sharing the same mass and charge. Furthermore, it also indicates that due to the non-linear complex scalar field, the mass of the scalarized charged black hole could smaller than its total charge.

\begin{figure}[htpb]
  \centering
  \includegraphics[width=0.3\textwidth]{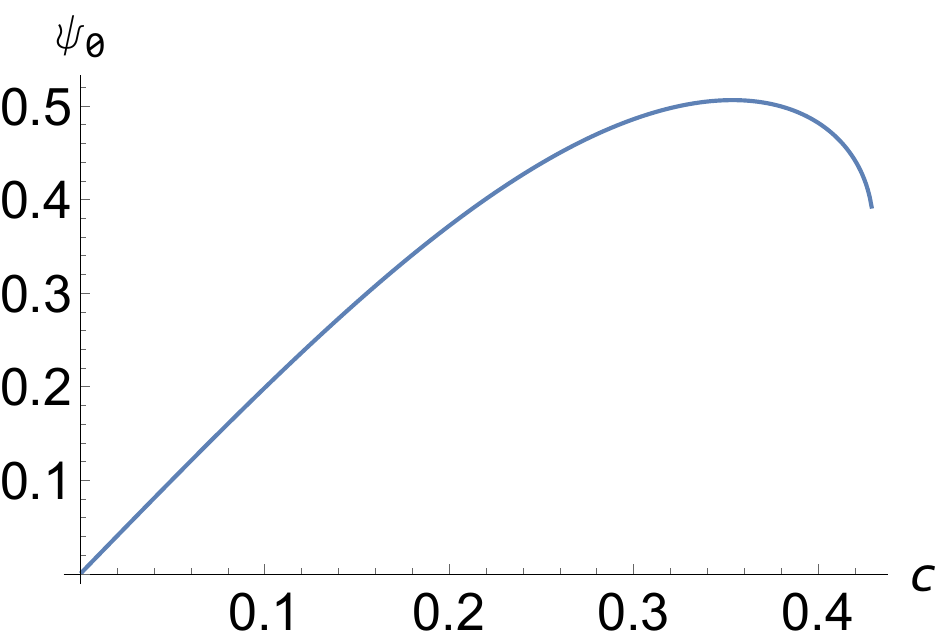}
  \includegraphics[width=0.3\textwidth]{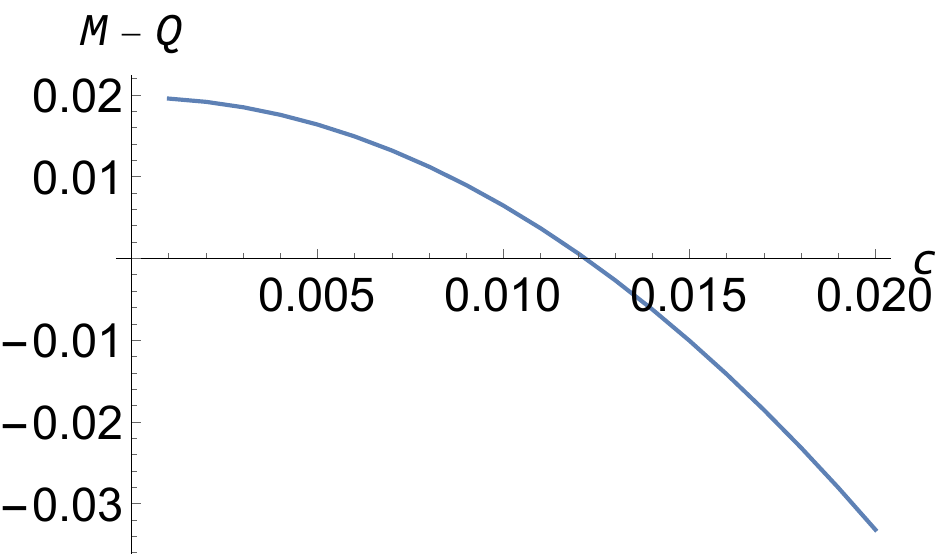}
  \includegraphics[width=0.3\textwidth]{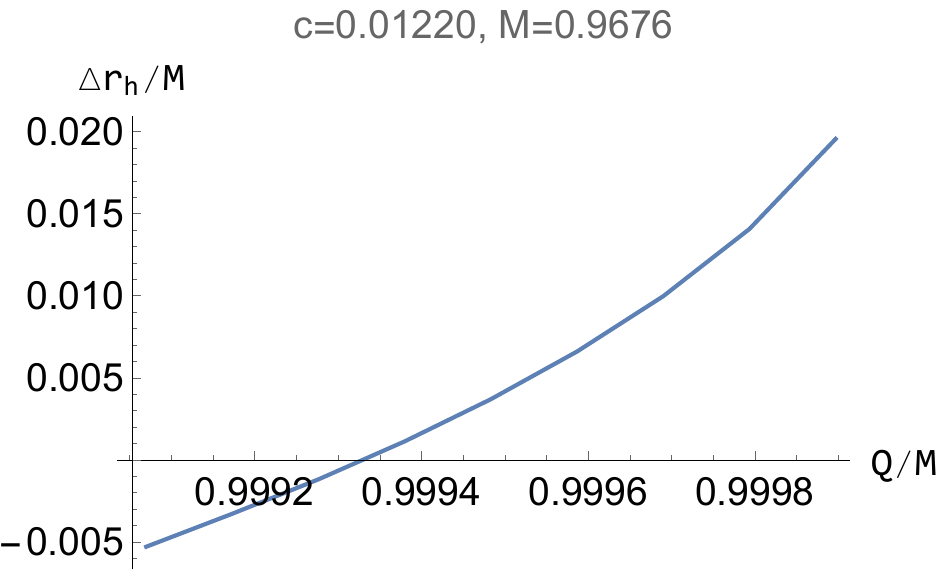}
 \caption{\textbf{Top}: Relationship between $c$ and $\psi_0$ based on our numerical solutions. \textbf{Middle}: A typical example about the relation between $M-Q$ and $c$. \textbf{Bottom}: The relation between $\Delta r_h/M$ and $Q/M$ with fixing $M=0.9676$. }\label{fig2}
\end{figure}

In following we search the parameter region in which the mass of the scalarized charged black hole is larger than its charge using shooting method where both scalarized black hole (if exists) and RN black hole are solutions of Eq.~\eqref{eqscalar1}. In this paper, we mainly adopt the shooting method in numerics for investigating the thermodynamical stability associated with the scalarized charged black hole.
%Generally, the shooting method can be written as the following simply form,
%%
%\begin{equation} \label{map2}
%\boxed{\{\text{Constrains} \}  \xrightarrow[\text{\{Initial Value\}}]{\text{Shooting}} \{\text{Solution}\}. }
%\end{equation}
%%
%Specially, to obtain a series of scalarized charge black hole solutions, we have
%%
%\begin{eqnarray}
%&& \{\psi_+ =0  \}  \xrightarrow[\{c_{i-1}, \psi_{0_{i-1}} \}]{\text{Shooting}} \{c_{i}, \psi_{0_{i}} \}, \quad i>0 \,  , \cr
%&& \{\psi_+ =0  \}  \xrightarrow[\{c_{i+1}, \psi_{0_{i+1}} \}]{\text{Shooting}} \{c_{i}, \psi_{0_{i}} \}, \quad i<0 \, .
%\end{eqnarray}
%%
We give a plot in $(c,\psi_{0})$ plane (See also the first plot of Fig.~\ref{fig2}), where every point on the curve denotes a numerical solution of charged black hole with scalar hair. Remarkably, from this plot one can easily see that $\psi_0$ will vanish when $c \ll 1$, implying that the scalarized charged black hole will reduce to RN black hole. Said another way, our result is consistent with the case of the probe limit in which the action Eq.~\eqref{action1} will reduce to Einstein-Maxwell theory after rescaling $\psi  \to c \psi$ then taking $c \to 0$.

Recall the Eq.~\eqref{MQ}, one can pick out the ADM mass $M$ and total charge $Q$ of every scalarized charged black hole denoted by $\{c_{i}, \psi_{0_{i}}\}$. Therefore, we present a plot on $(c, M-Q)$ plane in points and find $M-Q$ exist a zero point around $c \in [0.00101, 0.02]$, (See also the second plot of Fig.~\ref{fig2}). Using both the interpolation method and the shooting method, we obtain a series of  numerical scalarized charged black hole solution with $M=Q \approx 0.9676$, denoted by $\{c \approx 0.01220, \psi_0 \approx 0.02473 \}$, which is actually a good seed numerical solution for for investigating the relation between $\Delta r_h$ and $Q$ practically.
%Using shooting method that
%%
%\begin{eqnarray}
%&&\{M=M_i = 0.9676, Q=Q_i ,\psi_+=0 \}   \xrightarrow[\{r_{h_{i-1}}, \phi_{1_{i-1}}, \psi_{0_{i-1}} \}]{\text{shooting}} \cr
%&&\{r_{h_{i}}, \phi_{1_{i}}, \psi_{0_{i}} \}\, ,
%\end{eqnarray}
%%
Furthermore, recall that we are working with the microcanonical ensemble, the radius of the corresponding RN black hole  $r_{h_{RN i}} $ reads $r_{h_{RN i}}=M+\sqrt{M^2-Q_i^2} $. In other words.  we established a relation between $\Delta r_{h_i} = r_{h_i}-r_{h_{RN i}}$ and $Q_{i}$ (See also the last plot in Fig.~\ref{fig2} and the Appendix.~\ref{micapp} for detail discussion.).

From $S \sim r_h^2$ and the plot presented in $(Q/M, \Delta r_h/M)$ plane, one can find that there does exist a small interval $Q/M \in [0.9993,1]$ in which the entropy of the scalarized charged black hole is larger than the RN black hole in microcaonical ensemble. To conclude, in microcanonical ensemble, the scalarized charged black hole with the mass-charge ratio lying on the interval $0.9993<Q/M<1$ will more stable than the RN black hole. We note the when $|Q|\sim M$, i.e. the corresponding RN black hole approaches to be extreme, the scalarized black hole has larger entropy. We have carefully checked and find that this is also true if the parameters are taken to be other different values. Thus, our numerical results imply that the near extreme RN black hole is not stable and will be spontaneously scalarized via a first order phase transition. Remarkably, when $Q>M$, there still exists scalarized charged black hole solutions but the corresponding RN black hole does not.

\subsubsection{New versions for Penrose-Gibbons conjecture}
In fact our numerical result does also definitely give a negative answer to a long-standing conjecture named Penrose-Gibbons conjecture. It was conjectured that, for all asymptotically flat black holes, the area of horizon $A_h$, the total mass $M$ and the total charge $Q$ will satisfy following inequality~\cite{Mars:2009cj}
\begin{equation}\label{penrgibbs}
  M\geq\sqrt{\frac{A_h}{16\pi}}+Q^2\sqrt{\frac{\pi}{A_h}}\,
\end{equation}
or a weaker version
\begin{equation}\label{penrgibbsb}
  \sqrt{\frac{A_h}{16\pi}}\leq\frac12\left(M+\sqrt{M^2-Q^2}\right)\,
\end{equation}
if the weak energy condition is satisfied. The saturation will appear only in RN black holes. pThese two inequalities are charged generalization of following Penrose inequality
\begin{equation}\label{penrose1}
  M\geq\sqrt{\frac{A_h}{16\pi}}
\end{equation}
Eq.~\eqref{penrose1} has been proven in general static case by a few of different methods~\cite{Mars:2009cj}. However, the proofs of charged generalizations~\eqref{penrgibbs} and \eqref{penrgibbsb} are still open. In spherical case, they reduce to
\begin{equation}\label{penrgibbs2}
  M\geq\frac{r_h}2+\frac{Q^2}{2r_h}\,
\end{equation}
and
\begin{equation}\label{penrgibbs3}
  r_h\leq\left(M+\sqrt{M^2-Q^2}\right)\,.
\end{equation}
and the weak energy condition reduces into the requirement of $T_{00}\geq0$ outside event horizon.  In our model, recall the energy momentum tensor given in Eq.~\eqref{Energy}, the $tt$ component of energy momentum tensor in spherical anstanz Eq.~\eqref{metric0} reads
\begin{equation}
T_{tt}=\frac{1}{2}q^2\phi^2\psi^2+\frac{1}{2}\te^{-\chi}f V(\phi)+\frac{1}{4}f(r)\phi'^2+\frac{1}{2}f^2\psi^2 \, .
\end{equation}
For a semi-definite non-linear potential $V(\phi)$, it can obviously to observe that the weak energy condition $T_{00} \geq 0$ is always satisfied. In addition, The inequalities~\eqref{penrgibbs} and \eqref{penrgibbsb} are two generalization of Penrose inequality in charged case. Though the Penrose inequality in static case (which is called Riemannian-Penrose inequality) has several proofs, the strength version ~\eqref{penrgibbs} has not been prove even in static spherical case. A serval proofs have been obtained by assuming that outside the black hole there are no charge current sources i.e. in electrovacuum and horizon is connected, see Refs.~\cite{Malec:1994sy,Hayward:1998jj,Gibbons:1998zr,Khuri2013}. For inequality~\eqref{penrgibbs}, a couterexample was found when the horizon is not connected~\cite{Weinstein2005}. Though the proof of inequality \eqref{penrgibbsb} has not been obtained yet in general case, as far as we know, no counterexample of weaker version~\eqref{penrgibbsb} was reported. Base on the inequality of arithmetic and geometric means $\frac{a+b}{2}\geq \sqrt{ab}$, a natural deduction of Eq.~\eqref{penrgibbs2} is $M \geq Q$, satisfying in RN black hole. For  inequality~\eqref{penrgibbs3} to make sense it is necessary that the spacetime should satisfies $M\geq |Q|$.

However, our numerical results offer a counterexample for both inequalities~\eqref{penrgibbs2} and \eqref{penrgibbs3}. Therefore, even in spherically symmetric case which has only one connected horizon, the inequalities~\eqref{penrgibbs} and \eqref{penrgibbsb} can still be broken.

\begin{figure}[htpb]
  \centering
  \includegraphics[width=0.3\textwidth]{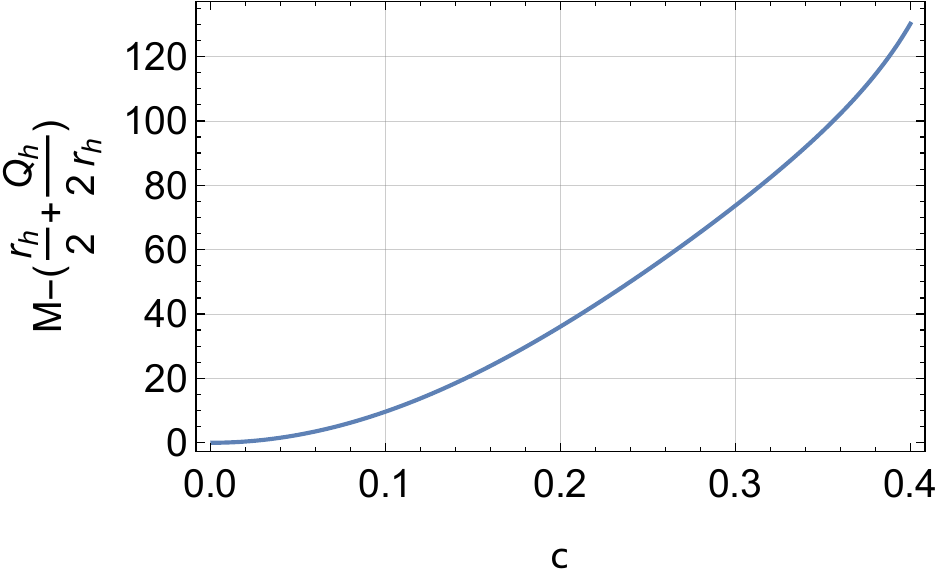}
  \includegraphics[width=0.3\textwidth]{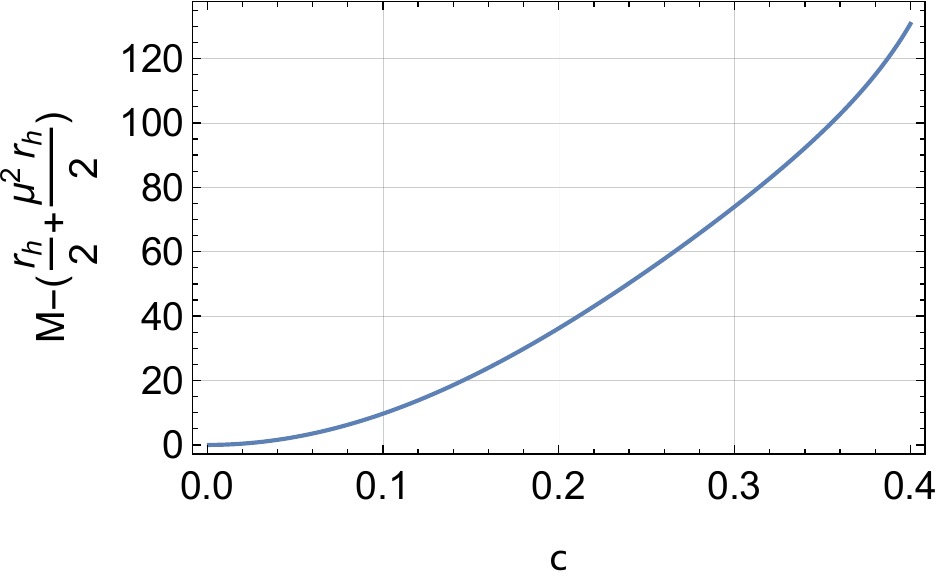}
 \caption{A numerical check on the inequalities shown in Eq.~\eqref{penrgibbs5}. Here we take $\{m=0.01, q=1.1m, r_h=1, \phi_1=0.8\}$.}\label{fig6}
\end{figure}

We note that inequalities~\eqref{penrgibbs} is not the only natural generalization of original Penrose inequality. Here we offer two new generalizations. One natural generalization of the Penrose inequality in scalarized charged black hole is that interpreting $Q$ as the charge enclosed with event horizon $Q_h$, which is identical in RN black hole due to the charge conservation law. The other natural generalization is that we use chemical potential $\mu$ to replace the role of charge $Q$. Therefore, we have two new versions of the generalized Penrose inequality,
\begin{equation}\label{penrgibbs4}
  M\geq\sqrt{\frac{A_h}{16\pi}}+Q_h^2\sqrt{\frac{\pi}{A_h}}\, , \quad M\geq\sqrt{\frac{A_h}{16\pi}}+\mu^2 r_h^2 \sqrt{\frac{\pi}{A_h}}.
\end{equation}
In spherical case, Eq.~\eqref{penrgibbs4} will reduce to
\begin{equation}\label{penrgibbs5}
  M\geq\frac{r_h}2+\frac{Q_h^2}{2r_h}\,, \quad  M\geq\frac{r_h}2+\frac{\mu^2 r_h}{2}\, .
\end{equation}
Here we only numerically verify these two inequalities Eq.~\eqref{penrgibbs5} respectively. Recall the class of scalarized charged black hole solution denoted by $\{c, \psi_0 \}$, we work with the $\left(c, M-\left(\frac{r_h}{2}+\frac{Q_h^2}{2r_h}\right)\right)$ plane and $\left(c, M-\left(\frac{r_h}{2}+\frac{\mu^2 r_h}{2 }\right)\right)$ respectively, where $Q_h= r_h^2 \te^{\chi(r_h)}\phi(r_h)$. From the plots given in Fig.~\ref{fig6}, we find that $M>\frac{r_h}{2}+\frac{Q_h^2}{2 r_h}$ and $M>\frac{r_h}{2}+\frac{\mu_h^2 r_h}{2}$ always hold as the growth of $c$, implying that the two possible generalization of the Penrose inequality we imposed hold in spherical case. Furthermore, it is an intriguing topic to proof Eq.~\eqref{penrgibbs3} in general scalarized charged black hole and we will leave it as future work.
\subsection{Canonical Ensemble and Grand Canonical Ensemble}
In this section, we turn to investigate the thermodynamic stability on scalarized charged black hole in both canonical ensemble and grand canonical ensemble, by means of  a series of specific numerical solution as an example. In canonical ensemble, the associated thermodynamic potential is the Helmholtz Free Energy $F(T,Q)$ with respect to the temperature $T$ and the total charge $Q$. The thermodynamics of the grand canonical ensemble can be described by the Gibbs Free Energy $G(T,\mu)$, in which the thermodynamic variables are the temperature $T$ and the chemical potential $\mu$ respectively. In both these two ensembles, the real physical process will be towards the direction increasing $F(T,Q)$ or $G(T,\mu)$. Thus, the RN black hole will spontaneously scalarize if the hairy black hole has smaller free energy.

Following the procedure developed by \cite{Gibbons:1976ue,Caldarelli:1999xj}, one can read off the free energy from the on-shell Euclidean action, namely $Z=\te^{-S_{\text{E}_{\text{on-shell}}}}$ and
\begin{equation}
F:=-\frac{1}{\beta}\ln Z|_{\text{canonical}}\,, \quad G:=-\frac{1}{\beta}\ln Z|_{\text{grand canonical}}\, ,
\end{equation}
where $Z$ is the thermodynamic partition function and $\beta=\frac{1}{T}$ is the inverse temperature. We start with the Euclidean action associated with \eqref{action1}
\begin{equation}\label{Eaction1}
S_E=S_{\text{bulk}}+S_{\text{surf}}+S_{\text{ct}} \, ,
\end{equation}
where
\begin{eqnarray}
&&S_{\text{bulk}}:=-\frac1{16\pi}\int \td^4 x \sqrt{g_E} \cr
&& \left(R-F_{\mu\nu}F^{\mu\nu}-(D_\mu\psi)^\dagger(D^\mu\psi)-W(|\psi|^2)\right)\, , \\
~ \cr
&&S_{\text{surf}} := -\frac{1}{8\pi}\int \td^3 x \sqrt{h_E}K \, , \\
~ \cr
&&S_{\text{ct}} := \frac{1}{8\pi}\int\td^3 x \sqrt{h_{\rm E}}K_0 -\frac{\alpha}{4\pi}\int \td^3 x \sqrt{h_{\rm E}} n_\nu F^{\mu\nu}A_\nu \, .
\end{eqnarray}
$S_{\text{surf}}$ and $S_{\text{ct}}$ denote the Gibbons-Hawking term and the counter terms respectively. Explicitly, $h_E$ and $K$ denote induce metric and the intrinsic curvature on arbitrary boundary $\partial M$. In addition, the first term of $S_{\text{ct}}$, $K_0 = \frac{2}{r}$, can remove the infinity arising from the spherical coordinates, while the second term with $\alpha$ is introduced for removing the boundary term relevant to the  electromagnetic field in the case of canonical ensemble. In other words, $\alpha=1,0$ denotes the case of canonical ensemble and grand-canonical ensemble respectively. In following we will present more illustration about the $\alpha$ term.

In the grand canonical ensemble, performing variation on the Euclidean action Eq.~\eqref{Eaction1} with respect to the fields gives
\begin{eqnarray}
\delta S  = &&-\frac{1}{16\pi} \int_{M} \td^4 x  \sqrt{g_E} \left (E_g^{\mu\nu}\delta g_{\mu\nu} + E_{\psi}\delta \psi \right) \cr
~\cr
&& -\frac{1}{4\pi}\int_{M} \td^4 x \sqrt{g_E} E_A^{\mu}\delta A_\mu   \cr
~\cr
&& + \frac{1}{4\pi}\int_{\partial M} \td^3 x \sqrt{h_E}n_\mu F^{\mu\nu}\delta A_\nu \, ,
\end{eqnarray}
where $n^\mu$ denotes the unit normal vector orthogonal to the boundary $\partial M$, $E_g^{\mu\nu}, E_\psi, E_A^\mu$ are the E.O.M given in Eq.~\eqref{eom}. For well-posed variation principle, one must impose an appropriate boundary condition. In general, $\delta A_{\mu}=0$ on $\partial M$ is imposed, which is appropriate for grand canonical ensemble with fixing chemical potential $\mu$. However, in the case of canonical ensemble with fixing the charge $Q$ on $\partial M$ as the boundary condition, the second term of $S_{\text{ct}}$ must be under consideration, namely $\alpha=1$. Moreover, the well-posed variation principle requires the boundary condition $\delta (n_\mu F^{\mu\nu})=0$ \cite{Hawking:1995ap,Caldarelli:1999xj}.

In the case that the scalar field is zero, the background is a RN black hole, of which the metric reads
\begin{eqnarray} \label{RNback}
&&\td s_{\text{RN}}^2= -f_{\text{RN}}(r)\td t^2+\frac{\td r^2}{f_{\text{RN}}(r)}+r^2 \td \Omega^2 \, , \cr
~\cr
&&f_{\text{RN}}(r)=1-\frac{2M}{r}+\frac{Q^2}{r^2}\, .
\end{eqnarray}
Plugging the RN black hole background Eq.~\eqref{RNback} into $S_{\text{E}}$ and setting $r \to \infty$ as the boundary, we find
\begin{equation}\label{EacRN}
S_{\text{E}_{\text{RN}}}=\frac{\beta}{2}(M \pm Q\Phi_{r_+})\, ,
\end{equation}
in which $\pm$ denotes the canonical ensemble case and the grand canonical ensemble case respectively. Then the Helmholtz free energy $F(T,Q)$ and the Gibbs free energy $G(\mu,T)$ of RN black hole can directly read
\begin{eqnarray}\label{Frn}
&&F_{\text{RN}}(T,Q)=\frac{1}{2}\left(M(T,Q) - \mu (T,Q) Q\right) \, , \\
~\cr
&&G_{\text{RN}}=\frac{1}{2}\left(M(\mu,T) + \mu Q(\mu,T)  \right)\, ,
\end{eqnarray}
in which the chemical potential $\mu= \frac{Q}{r_+} $ is interpreted as the electric potential at the infinity. In the canonical ensemble, the ADM mass of the RN black hole $M_{\rm RN}$ can be solved by the following relation
\begin{eqnarray} \label{rnTQ}
&& M_{\rm RN}=\frac{1}{2}\left(r_{h_{\rm RN}} + r_-\right) \, , \quad Q_{\rm RN}^2 =r_{h_{\rm RN}} r_- \, , \cr
 ~\cr
&& T_{\rm RN}=\frac{r_{h_{\rm RN}} - r_-}{4\pi r_{h_{\rm RN}}^2}\, ,\quad \mu=\frac{Q}{r_{h_{\rm RN}}}\, .
\end{eqnarray}
where $r_-$ is the inner horizon of the RN black hole. However, the analytical expression of $M_{\rm RN}(T_{\rm RN},Q_{\rm RN})$ is too complicated to present since there are three real or complex roots. In practice, numerically selecting the real root of $M_{\rm RN}(T_{\rm RN}, Q_{\rm RN})$ with $M_{\rm RN}>\frac{r_{h_{\rm RN}}}{2}$, we then read off $F_{\rm RN}(T_{\rm RN},Q_{\rm RN})$. As to in grand canonical ensemble, the Eq.~\eqref{rnTQ} gives
\begin{equation}\label{rnmuT}
M_{\rm RN}=\frac{1- \mu_{\rm RN}^4}{8 \pi T_{\rm RN}} \, , \quad Q_{\rm RN}=\frac{\mu_{\rm RN}(\mu_{\rm RN}^2-1)}{4\pi T_{\rm RN}}\, .
\end{equation}
We thus read off the value of $G_{\rm RN}(\mu,T_{\rm RN})$ from Eq.~\eqref{Frn} as well.

\subsubsection{The case of probe limit $c \ll 1$}
Now we turn to consider the case of the scalarized charged black hole. We firstly present a proof that, in the probe limit $c \to 0$, the RN black hole will be more stable than the scalarized charged black hole in both canonical ensemble and the grand canonical ensemble. Firstly, rescaling $\psi$, namely $\tilde{\psi} \to c \psi$, gives
\begin{equation}\label{Eaction2}
 S_\text{E}=S_{\text{E}_{\text{EM}}}+\frac{\beta c^2}{16\pi}\int \td^4 x \sqrt{g_{\text{E}}} (D^ \mu \tilde{\psi})^{\dag}(D_\mu \tilde{\psi})+m^2 \log(1+|\tilde{\psi}|^2) \, ,
\end{equation}
where $S_{\text{E}_{\text{EM}}}$ denotes the Euclidean action contributed by Einstein-Maxwell theory.  At the limit $c \to 0$, it is clear that the solution will be a RN black hole with the metric $g_{\mu\nu}^{\text{RN}}$ and gauge potential $A_{\mu}^{\text{RN}}$. When $c\neq 0$ but $c\ll q$, one can treat the contribution of scalar field as a perturbation of order $\mathcal{O}(c^2)$. Let us assume that the metric and gauge potential become
\begin{equation}\label{metricas1}
  g_{\mu\nu}=g_{\mu\nu}^{\text{RN}}+c^2g_{\mu\nu}^{(1)},~~A_\mu=A_{\mu}^{\text{RN}}+c^2A_\mu^{(1)},~~~c^2\ll1\,.
\end{equation}
Upon the $\mathcal{O}(c^2)$ order, we find the Euclidean action contributed by Einstein-Maxwell theory reads
\begin{eqnarray}\label{actems1}
&& S_{\text{E}_{\text{EM}}}=S_{\text{E}_{\rm RN}}  + c^2\int_{M}\sqrt{g_E}\td^4x\left[(G_{\mu\nu}|_{g_{\mu\nu}=g_{\mu\nu}^{\text{RN}}})g_{\mu\nu}^{(1)} \right. \cr
~\cr
&&+ \left.(\nabla_{\mu}F^{\mu\nu}|_{A_\mu=A_{\mu}^{\text{RN}}})A_\nu^{(1)}\right]+\mathcal{O}(c^4)\,,
\end{eqnarray}
where $S_{\text{E}_{\rm RN}}$ is the on-shell action in RN background given by
\begin{equation}
G_{\mu\nu}|_{g_{\mu\nu}=g_{\mu\nu}^{\text{RN}}}=\nabla_{\mu}F^{\mu\nu}|_{A_\mu=A_{\mu}^{\text{RN}}}=0 \, .
\end{equation}
Therefore, we find the $S_{\text{E}_{\text{EM}}}=S_{\text{E}_{\rm RN}}+\mathcal{O}(c^4)$, which implies upon the leading order of $c$, the on-shell Euclidean action of Eq.~\eqref{Eaction2} can be regard as
\begin{equation} \label{Eaction3}
S_{\text{E}_{\text{on-shell}}}=S_{\text{E}_{ \rm RN}}+S_{\text{E}_\text{scalar}} \,
\end{equation}
with
\begin{equation} \label{Eaction3b}
S_{\text{E}_\text{scalar}}=\frac{\beta c^2}{16\pi}\int \td^4 x \sqrt{g_{\text{E}}}[ ( \bar{D}^ \mu \tilde{\psi})^{\dag}(\bar{D}^{\mu} \tilde{\psi})+m^2 \log(1+|\tilde{\psi}|^2)] \, ,
\end{equation}
where $\bar{D}_\mu $ denotes the covariant derivative under the RN black hole background~\eqref{RNback}.

Plugging the RN black hole background Eq.~\eqref{RNback} into Eq.~\eqref{Eaction3b},  we have
\begin{eqnarray} \label{Scaos}
&&S_{\text{E}_\text{scalar}}= \frac{\beta c^2}{4}\int^{\infty}_{r_{h_{\rm RN}}} \td r ~r^2 \cr
~\cr
&&\left(f_{\text{RN}} \tilde{\psi'}^2 + \frac{q^2 \phi_{\rm RN}^{2} \tilde{\psi}^2}{f_{\text{RN}}} +m^2 \log(1+|\tilde{\psi}|^2)\right), \,
\end{eqnarray}
where the prime means the derivative with respect to $r$. It needs to note $\phi_{\rm RN}^{2}=-Q^2/r^2\leq0$, since $\phi_{\rm RN}$ denotes the electric potential of the RN black hole in Euclidean spacetime, namely $A^{\rm RN}_{\mu}= \phi_{\rm RN}(\td\tau)_{\mu}=\frac{i Q}{r}$ (here $\tau$ is  Euclidean time). This leads that the sign in the integration is undefined. Note that the equation of motion associated with the scalar field in RN background
\begin{equation}
\bar{D}^\mu \bar{D}_{\mu}\tilde{\psi} - \frac{m^2}{1+|\tilde{\psi|}^2}\psi=0,
\end{equation}
where explicitly given
\begin{equation}
f_{\rm RN}\tilde{\psi}''+f_{\rm RN}'\tilde{\psi}'+\frac{2f_{\rm RN}}{r}\tilde{\psi}'-\frac{m^2}{1+|\tilde{\psi}|^2}\tilde{\psi}- \frac{q^2 \phi_{\rm RN}^2}{f}\tilde{\psi}=0
\end{equation}
Then the Eq.~\eqref{Scaos} gives
\begin{eqnarray}\label{Scaos2}
&&S_{\text{E}_\text{scalar}}=\frac{\beta c^2}{4}\left( \left(r^2 \tilde{\psi} \tilde{\psi}' f_{\rm RN}\right)|^{\infty}_{r_{h_{\rm RN}}} \right. \cr
~\cr
&&+ \left.\int^{\infty}_{r_{h_{\rm RN}}} r^2
m^2 \left(\log(1+|\tilde{\psi}|^2)-\frac{\tilde{\psi}^2}{1+|\tilde{\psi}|^2}\right)\td r  \right)\, .
\end{eqnarray}
Given appropriate boundary condition $\psi (\infty)=0$ and $f_{\rm RN}|_{r=r_{h_{\rm RN}}}=0$, the first term of Eq.~\eqref{Scaos2} will vanish. To proceed, we consider the second term of Eq.~\eqref{Scaos2}. One can construct a auxiliary function $t(x)$ as
\begin{equation}
t(x)=\log(1+x)-\frac{x}{1+x}\, , \quad  \frac{\td t(x)}{\td x}=\frac{x}{(1+x)^2} .
\end{equation}
It is easy to verify that for $x>0$, $ \frac{\td t(x)}{\td x}>0$ always hold, indicating $t(x)$ is a monotone increasing function where the minimum value is lying on $x=0, t(0)=0$. Since $|\psi|^2$ is positive for any $r$, the integrand of the second term of Eq.~\eqref{Scaos2} is always positive. Therefore, we have proved
\begin{equation}\label{frees1}
S_{\rm{E}_\text{scalar}}>0.
\end{equation}
Furthermore, according to $F=\frac{1}{\beta} S_{\text{E}},$ and $G=\frac{1}{\beta} S_{\text{E}}$, we have
\begin{eqnarray}
&\Delta F = F-F_{RN}=F_{\text{scalar}}>0 \, , \\
&\Delta G = G-F_{RN}=G_{\text{scalar}}>0\, .
\end{eqnarray}
To conclude, base on our demonstration above, the contribution of the complex scalar field to the $F(T, Q)$ and $G(T, \mu)$ will be always positive. According to the stability requirement that the smaller free energy indicates the more stable of the black hole, we thus  prove that in probe limit, the RN black hole is more stable than the scalarized charged black hole in both canonical ensemble and grand canonical ensemble.

\subsubsection{The case of general $c$: Canonical Ensemble}
When $c$ is not infinitesimal, the higher order terms of $c$ play role and above proof is broken. In the following, we consider the Euclidean action Eq.~\eqref{Eaction1} in general. As there is not analytical solution for scalarized black hole, we can only compute the free energy numerically. In order to do that, let us first some useful formulas, which can simplify the numerical computation of free energy. Performing the Wick rotation, the Euclidean line element gives
\begin{equation}\label{metric0E}
  \td s^2=f(r)\te^{-\chi(r)}\td \tau^2+\frac{\td r^2}{f(r)}+r^2\td\Omega^2
\end{equation}
and
\begin{equation}\label{matters1E}
  A_\mu=i\phi(r)(\td \tau)_\mu,~~\psi=\psi(r)\,.
\end{equation}
Note that the trick given in \cite{Hartnoll:2008kx}, we also find the following relation between the on shell Lagrangian and the $\theta\theta$ component of the energy momentum tensor,
\begin{equation}
2T^{\theta}{}_{\theta}={\cal L_{\text{on-shell}}}-R.
\end{equation}
Consider the equation of motion given in Eq.~\eqref{eom}, we arrive
\begin{equation}
{\cal L}_{\rm on-shell}=-{G^t}_t-{G^r}_r=\frac2{r^2}[(rf)'+r\chi'f-1] \, ,
\end{equation}
where $G_{\mu\nu}$ is the Einstein tensor. Recall the Euclidean on-shell action of RN black hole Eq.~\eqref{EacRN}, we obtain the Euclidean on-shell action of Eq.~\eqref{Eaction1},
\begin{equation}
S_{\text{E}_{\text{on-shell}}}=\frac{\beta M}{2}+\frac{\beta}{2} \int^\infty_{r_h} \td r[(rf\te^{-\chi/2})'-\te^{-\chi/2}]+\alpha \mu Q \, .
\end{equation}
After performing integration by parts, we respectively obtain the Gibbs free energy in grand canonical ensemble
\begin{equation}\label{freeF2}
G(\mu,T)=\frac{r_h}{2}-\frac{M}2+\frac12\int_{r_h}^\infty(1-\te^{-\chi/2})\td r \, ,
\end{equation}
and the Helmholtz free energy in canonical ensemble
\begin{equation}\label{freeF1}
F(Q,T)=\frac{r_h}2-\frac{M}2+\frac12\int_{r_h}^\infty(1-\te^{-\chi/2})\td r + \mu Q \, .
\end{equation}
Given a scalarized charged black hole $\{c_i,\psi_{0_i} \}$, the Hawking temperature and the chemical potential read
\begin{equation} \label{Tmu}
T=\frac{f'(r_h)\te^{-\chi(r_h)/2}}{4\pi}\,  , \quad \mu = [\phi(r) + r \phi'(r)]|_{r \to \infty}.
\end{equation}
%

\begin{figure}
  \centering
  \includegraphics[width=0.3\textwidth]{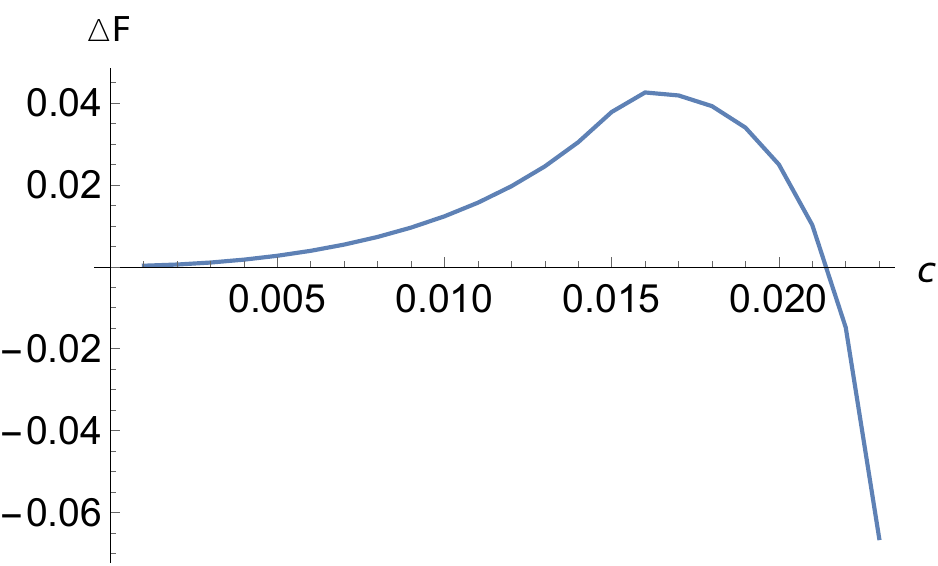}
  \includegraphics[width=0.3\textwidth]{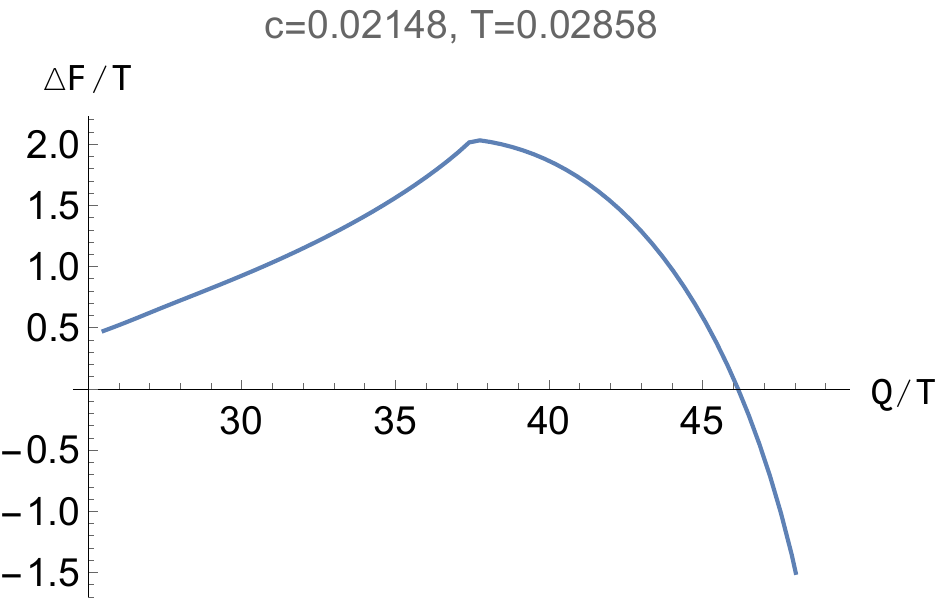}
  \includegraphics[width=0.3\textwidth]{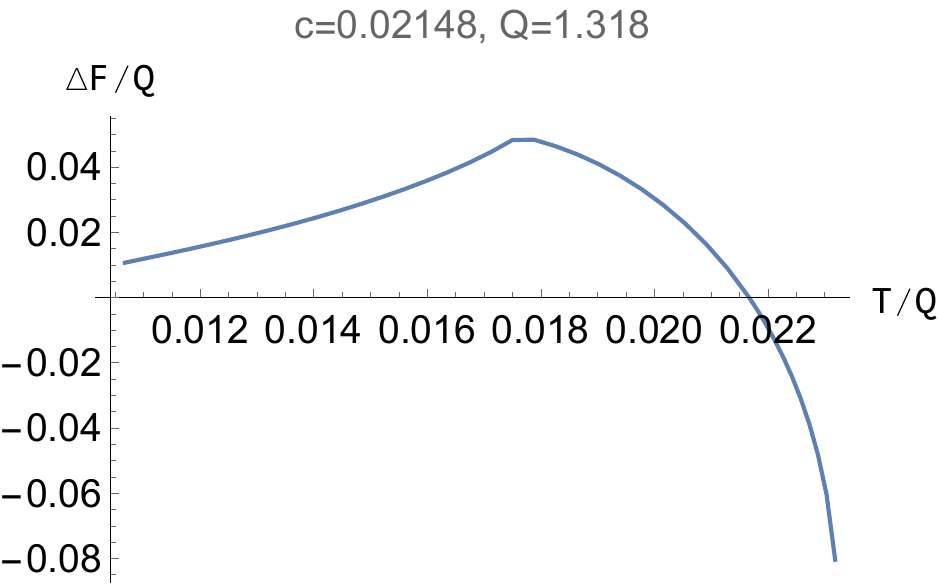}
 \caption{\textbf{Top}: A typical example about relationship between $\Delta F$ and $c$. Here we take $\{r_h=1, m=0.01, q=11m, \phi_1=0.8\}$. \textbf{Middle}: The relation between $Q/T$ and $\Delta F/T$ by fixing $T=0.02858$. \textbf{Bottom}: The relation between $T/Q$ and $\Delta F/ Q$ with fixing $Q=1.318$. }\label{fig3}
\end{figure}

%To numerically investigate the behaviour of $\Delta F = F - F_{\text{RN}}$ in canonical ensemble and  $\Delta G= G - G_{\text{RN}}$ in grand canonical ensemble, we still adopt the approach given in Sec.~\ref{micensem}.
In the case of canonical ensemble, we pick out the temperature $T$, total charge $Q$ and evaluate the helmholtz free energy $F(T,Q)$ of every scalarized charged black hole solution $\{c_i, \psi_{0_i}\}$ base on Eq.~\eqref{MQ},  Eq.~\eqref{Tmu} and Eq.~\eqref{freeF1}. Using shooting method,
%%
%\begin{eqnarray}
%&& \{T=T_i = 0.02858, Q=Q_i ,\psi_+=0 \}   \xrightarrow[\{r_{h_{i-1}}, \phi_{1_{i-1}}, \psi_{0_{i-1}} \}]{\text{shooting}} \cr
%&&\{r_{h_{i}}, \phi_{1_{i}}, \psi_{0_{i}} \}\, ,  \\
%~\cr
%&& \{T=T_i, Q=Q_i=1.318 ,\psi_+=0 \}   \xrightarrow[\{r_{h_{i-1}}, \phi_{1_{i-1}}, \psi_{0_{i-1}} \}]{\text{shooting}}  \cr
%&& \{r_{h_{i}}, \phi_{1_{i}}, \psi_{0_{i}} \}\, .
%\end{eqnarray}
%%
we work with dimensionless $(Q/T, \Delta F/T)$ with fixing temperature $T=T_0=0.02858$ and $(T/Q, \Delta F/Q)$ plane with fixing $Q=Q_0=1.318$ without losing generality. (See also the middle and the right plot in Fig.~\ref{fig3} and Appendix.~\ref{canapp} for detail discussion).

From the plot in $(Q/T, \Delta F/T)$ and $(T/Q, \Delta F/Q)$ plane we present in Fig.~\ref{fig3}, one can find that there also exist small intervals $Q/T \in [46.15,49.30] $ and $T/Q \in [0.02167,0.02318]$ in which the $F(T,Q)$ of sclarized charged black hole is smaller than the corresponding RN black hole in canonical ensemble. To summarize, this result indicates that in the case of canonical ensemble, the scalarized charged black hole is more stable than the corresponding RN black hole in the region $Q/T \in [46.15,49.30] $ and $T/Q \in [0.02167,0.02318]$. Remarkably, when $Q/T>49.30$ and $T/Q>0.02318$ region there does not exist the corresponding RN black hole.

\subsubsection{The case of general $c$: Grand Canonical Ensemble}
\begin{figure}[htpb]
  \centering
  \includegraphics[width=0.3\textwidth]{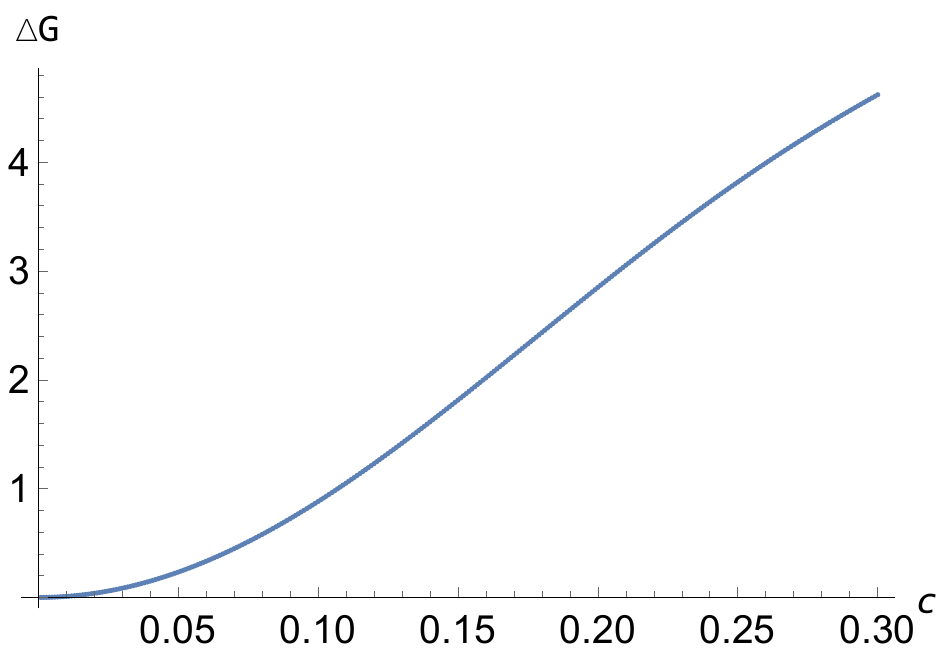}
  \includegraphics[width=0.3\textwidth]{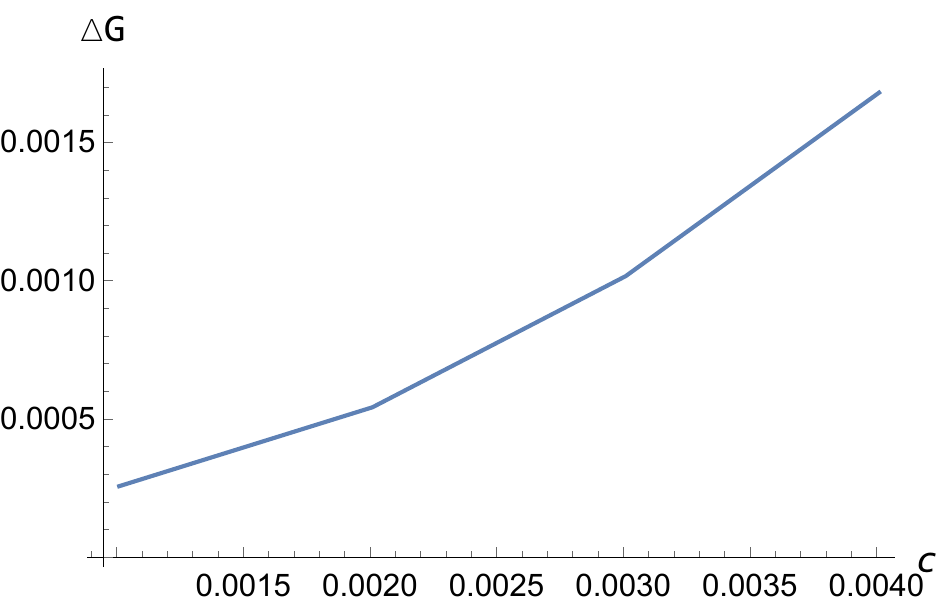}
 \caption{\textbf{Top}: A typical example about the relationship between $\Delta G$ and $c$. Here we take $c \in [0.001, 0.4]$.
 \textbf{Bottom}: Zooming in the left plot in the small $c$ region $c \in [0.001, 0.004]$. Here we take $\{r_h=1, m=0.01, q=11m, \phi_1=0.8\}$.}\label{cG}
\end{figure}

\begin{figure}[htpb]
  \centering
  \includegraphics[width=0.3\textwidth]{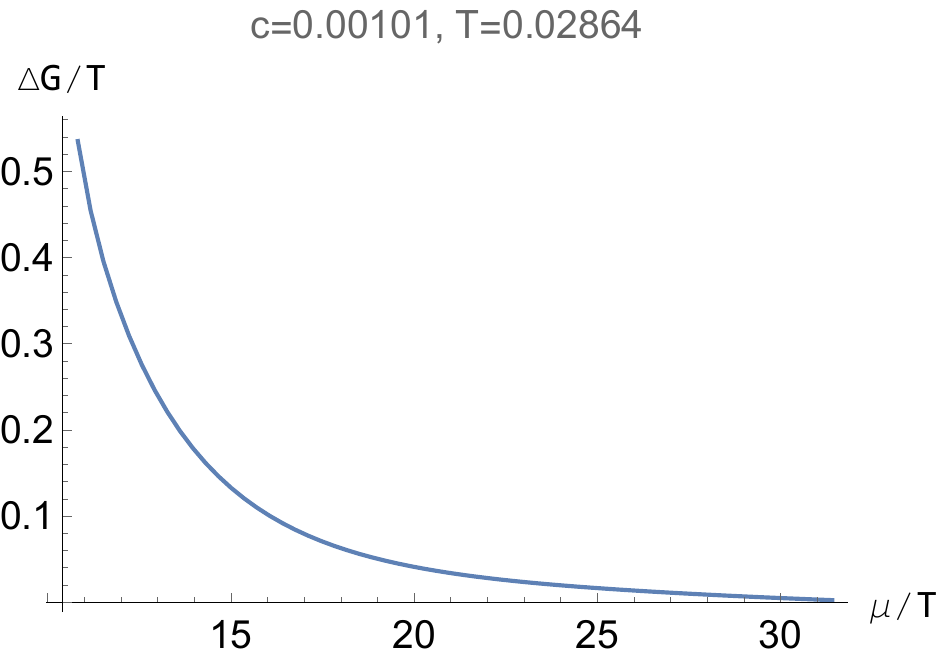}
  \includegraphics[width=0.3\textwidth]{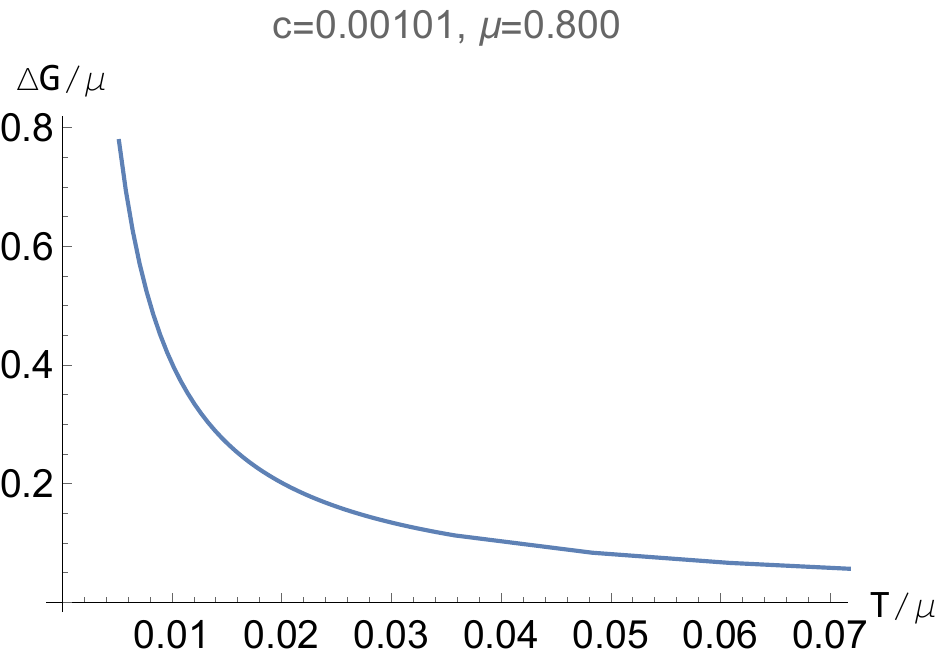}
 \caption{In the case of $c=0.00101$, \textbf{Top}:  A typical example about the relation between $\mu/T$ and $\Delta G/T$ with fixing $T=0.02864$. \textbf{Bottom}: The other typical example about the relation between $T/\mu $ and $\Delta G/\mu)$ with fixing $\mu=0.800$. }\label{GTmu}
\end{figure}

Finally, we turn to consider the grand canonical ensemble case where the thermodynamic variables is temperature $T$ and chemical potential $\mu$. Different from in the case of microcanonical ensemble and canonical ensemble, from Eq.~\eqref{rnmuT} one will find that given a numerical scalarized charged black hole solution with temperature $T_i$ and chemical chemistry $\mu_i$, there is always a corresponding RN black hole sharing the same $T_i$ and $\mu_i$. With this in mind, we pick out $\Delta G(T, \mu)= G(T, \mu)-G_{\rm RN} (T,\mu)$ and present a plot in $(c, \Delta G)$ plane (See also the plots in Fig.~\ref{cG}) from which one can observe that $\Delta G$ is always positive for any $c_{i}$.
The absence of appropriate seed solution indicates that the Gibbs free energy $G(T, \mu)$ of an arbitrary scalarized charged black hole solution might be always larger than the corresponding RN black hole in grand canonical ensemble. In following we take a small $c=0.00101$ and larger $c=0.1$ as example to study the behavior of $\Delta G$ related to the fixing temperature $T$ and chemical potential $\mu$. In this case, one can read off the $G_{\rm RN}(T, \mu)$ of the corresponding RN black hole with temperature $T_i$ and chemical potential $\mu_i$ from Eq.~\eqref{Frn} and Eq.~\eqref{rnTQ}. We then work with a dimensionless $(\mu/T, \Delta G/T)$ plane by fixing $T=0.02864$ and dimensionless $(T/\mu, \Delta G/ \mu)$ plane with fixing $\mu=0.800$ respectively, without losing generality as we mentioned above. We also offer two plots as well (See also the plots presented in Fig.~\ref{GTmu} and Appendix.~\ref{graapp} for detail).
These results show that the $G(T,\mu)$ of the scalarized charged black hole is always larger than the corresponding RN black hole in grand canonical ensemble when $c$ is small. Therefore it can be concluded that the RN black hole is more stable than scalarized charged black hole in grand canonical ensemble, which is consistent with our proof in probe limit.

\begin{figure}[htpb]
  \centering
  \includegraphics[width=0.3\textwidth]{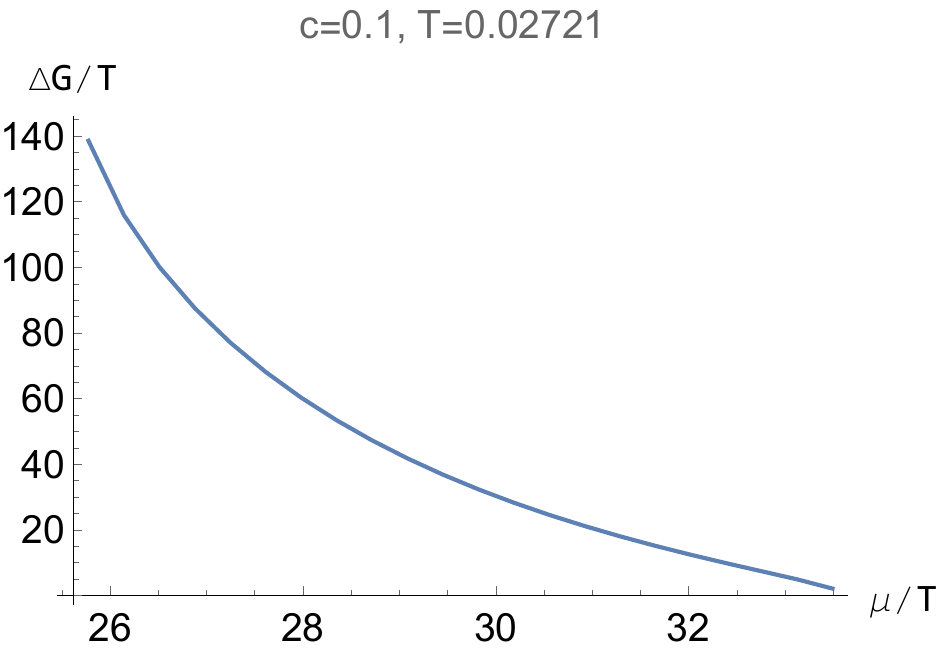}
  \includegraphics[width=0.3\textwidth]{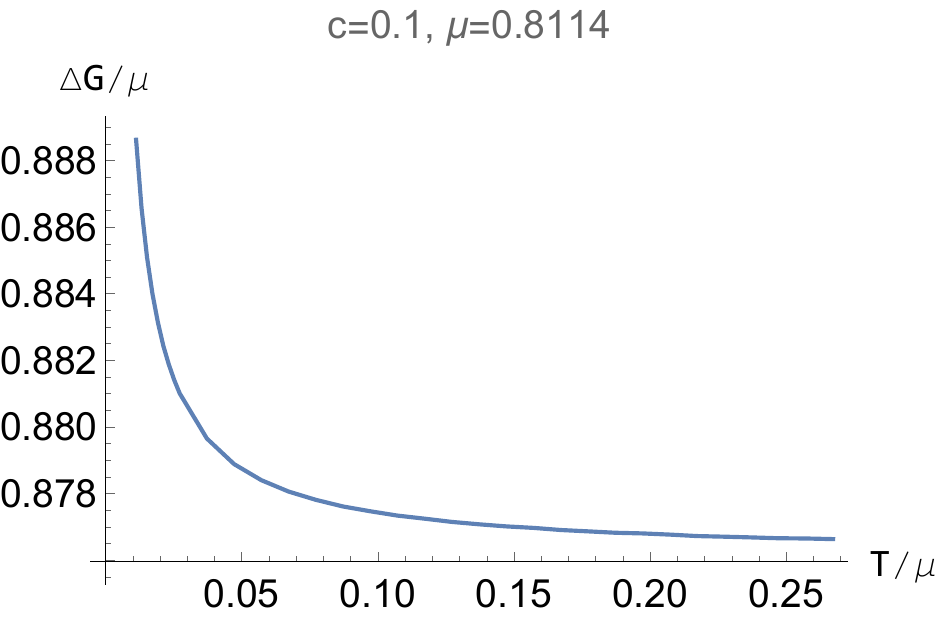}
 \caption{In the case of  $c=0.1$, \textbf{Top}: A typical figure about the relation between of $\mu$ and $\Delta G$ in dimensionless $(\mu/T, \Delta G/T)$ plane with fixing $T=0.02721$.  \textbf{Bottom}:  Another typical example about the relation between of $T$ and $\Delta G$ in dimensionless $(T/\mu, \Delta G/\mu)$ plane with fixing $\mu=0.8114$.}\label{GTmu2}
\end{figure}

As to Working with dimensionless $(\mu/T, \Delta G/T)$ plane with fixing $T=0.02721$ and dimensionless $(T/\mu, \Delta G/\mu)$ plane with fixing $\mu=0.8114$ respectively, we present a plot (See the plots of Fig.~\ref{GTmu2})
and observe that $\Delta G$ is also positive for any $T$ and $\mu$ from the plots, it can be concluded that for a larger $c$, the RN black hole is also more stable than the scalarized charged black hole in grand canonical ensemble. We have check carefully for other different parameters and find the same conclusion. Hence, we summarize that in thermodynamics, the RN black hole is more stable than the scalarized charge black hole in grand canonical ensemble.

\section{Kinetic Stability}\label{KietSta}
In this section, we turn to study the kinetic stability of the scalarized charged black hole solution, i.e. the stability against a small perturbation. At first, based on perturbation theory of black hole, we setup a general model to obtain one-dimension, Schr\"{o}dinger like radial function in frequency domain. We then investigate the validity of the shooting method numerically through analyzing numerical error, pointing out that the shooting method can catch the unstable modes efficiently, rather than stable modes. Based on the argument in \cite{Konoplya:2019hlu}, we adopt the WKB approximation method to calculate the stable mode of perturbative field since the damping mode can be given by means of the WKB approximation method.

Practically, considering a massless real perturbative scalar field, we obtain the linearized equation of motion associated with the scalar perturbation. Due to the static and spherical black hole background, the equation of motion of the scalar perturbation can be reduced to 1 dimension in frequency domain\cite{Chandrasekhar:1975zza}. Imposing an appropriate physical boundary condition, ingoing wave near the horizon and outgoing wave at the spatial infinity, resonance state of the scalar perturbation arise, picking out a class of complex frequency $\omega = \omega_{\rm R}+ i \omega_{\rm I}$ so called black hole quasi-normal modes (QNMs).  Moreover, the imaginary part of frequency $\omega_{\rm I}$ indicates energy dissipation at both the horizon and the spactial infinity. If $\omega_{\rm I}>0$, the scalar perturbation will grow exponentially, leading to the instability of the black hole background at least at linear level. If $\omega_{\rm I}<0$, the scalar field will not trigger on the unstable of spacetime for exponentially damp, and finally dissipative out rapidly.

Firstly, we begin with reducing the master equation of the scalar perturbation in radial equation in frequency domain. Given a spherical line element Eq.~\eqref{metric0}, the equation of motion of the perturbative scalar field gives
\begin{equation}\label{proeom}
\square \psi_2 =0.
\end{equation}
where $\square=\nabla^\mu \nabla_\mu$ is the d'Alembert operator in spherical line element background, Eq.~\eqref{metric0}. By adopting the separation variables method, we take the following anstaz of $\psi_2$ under the spherical background,
\begin{equation}
\psi_2=\te^{-i \omega t}R(r)Y_{lm}(\theta,\varphi),
\end{equation}
where $l$ and $m$ are the azimuthal quantum number and magnetic quantum number respectively, satisfying $l>m, m=0,1,2,...$. The radial equation reads
\begin{equation}\label{radical}
\Delta \frac{d}{dr}\Delta \frac{dR}{dr}+ U R =0,
\end{equation}
where
\begin{equation}
\Delta=r^2 f(r)\te^{-\frac{\chi}{2}}\, , \quad U=\frac{\omega^2 r^2}{f \te^{-\frac{\chi}{2}}}-l(l+1).
\end{equation}
By introducing the tortoise coordinate and a new radial equation $\tilde{R}$ as
\begin{equation}\label{tortoise}
\frac{dy}{dr}=\frac{r^2}{\Delta}\, , \quad R=r \tilde{R}\, ,
\end{equation}
the radial equation Eq.(\ref{radical}) can be written in the following standard wave function form

\begin{equation}\label{derad}
\frac{d^2 \tilde{R}}{dy^2}+\tilde{U}\tilde{R}=0 \, ,
\end{equation}
where
\begin{eqnarray}\label{UU}
\tilde{U}&=&\omega^2-V \cr
~\cr
&=&\omega^2-\frac{l(l+1)}{r(y)^2}f(y) \te^{-\chi(y)} \cr
~ \cr
&&-\frac{1}{2r(y)}\frac{d}{dr}\left(f(y)^2 \te^{-\chi(y)}\right)\, .
\end{eqnarray}
In Eq.~\eqref{UU}, we denotes $f(y), \chi(y)$ and $\chi(y)$ as a function with respect to $y$. From the definition of the tortoise coordinate, one can read $y \to -\infty$ corresponding to $r \to r_h$, while $y \to \infty$ corresponding to $r \to \infty$. To single out a series of QNMs, we impose the following boundary condition,
\begin{equation}\label{wavebound}
\tilde{R} \varpropto \left\{
\begin{aligned}
&\te^{-i \omega y } \quad \quad  &&y \to -\infty \\
&\te^{i \omega y}   \quad \quad  &&y \to \infty \, ,
\end{aligned}
\right.
\end{equation}
indicating that the probing scalar is pure going wave at the horizon and pure outgoing wave at the spatial infinity. In general, in order to investigate both the stable and unstable mode of perturbative field, one can solve QNMs using numerical method, popularly the shooting method. There are also serval effective approximative methods to calculate QNMs, for example, the WKB approximation method. It has been argued that the WKB approximation can only catch the stable modes with $\omega_{\rm I}<0$ even if the spectrum contains unstable modes~\cite{Konoplya:2019hlu}.  We will show that, on the contrary, the shooting method can only catch the unstable modes of $\omega_{\rm I}>0$ but cannot catch the stable modes. Thus, the combination of shooting method and the WKB approximation method can offer us a complete analyses on QNMs.

\subsection{Shooting Method and Analysis on Numerical Error}\label{numerr}
In this section, we analyze numerical error of shooting method and illustrate its validity on the calculation of the unstable modes. Let us first briefly explain how to use shooting method to find QNMs.

To setup the numerical procedure, we firstly interpret Eq.~\eqref{tortoise} and Eq.~\eqref{derad} as a set of ODE (one first-order differential equation as well as one second order differential equation), one thus need three boundary condition. In addition to Eq.\eqref{wavebound}, we need the last boundary condition associated to the radial coordinate $r$ pand tortoise coordinate $y$. we thus asymptotically expand near horizon. Specifically, in near horizon region, Eq.~\eqref{tortoise} gives
\begin{equation}\label{tortoise2}
\frac{\td y}{\td r}= \frac{1}{f_1 (r-r_h)}\, ,
\end{equation}
where $f_1$ is given by Eq.~\eqref{horbou1} and the equation of motion Eq.~\eqref{eqscalar1}. Interpreted the radial coordinate as a analytical function with respect to the tortoise coordinate $y$. The Eq.~\eqref{tortoise2} then can be solved that
\begin{equation}\label{wavebound0}
r=r_h+\te^{f_1 y}\, , \quad y \to -\infty \, .
\end{equation}
Therefore, we obtain a numerical solvable boundary-valued question with two differential equations
\begin{equation}\label{waveode}
\left\{
\begin{aligned}
&\frac{d^2 \tilde{R}}{dy^2}+\tilde{U}\tilde{R}=0 \, , \cr
&\frac{dr}{dy}=f(r(y))\, ,
\end{aligned}
\right.
\end{equation}
where $\tilde{U}$ has been given in Eq.~\eqref{UU}. To obtain a numerical eigenfunction of QNMs satisfying the boundary condition Eq.~\eqref{wavebound}, working with complex numerics, we primarily numerically perform integration on Eq.\eqref{waveode} with Eq.~\eqref{wavebound} and
\begin{equation}
\tilde{R} \varpropto \te^{- i \omega y} \, ,  \quad y \to \infty.
\end{equation}
In general, the numerical integration will give two branch solution in $y \to \infty$ region,
\begin{equation}
\tilde{R} \varpropto \te^{i \omega y} + {B} \te^{- i \omega y}\, , \quad y \to \infty.
\end{equation}
Furthermore, we have
\begin{equation}
{B} \varpropto \frac{\te^{i \omega y}}{2 \omega}\left(R \omega + i R'(y)\right) \, , \quad y \to \infty.
\end{equation}
The QNMs will then give
\begin{equation}
B \left(\omega_{\text{QNMs}}\right)=0\, .
\end{equation}
In practice, one can locate the QNMs by drawing a density figure related to $\frac{1}{B}$ in $\left(\omega_{\text{R}}, \omega_{\text{I}} \right)$ plane. Since the QNMs will give a vanish $R_{-}$, the location of QNMs in density figure can thus be observed as a bright spot in a density figure. Through observe the approximative region of the bright spot related to $\omega_{\text{R}}$ and $\omega_{\text{I}}$, denoted as $\omega_{\text{Initial}}=\omega_{\text{R}_{\text{Initial}}} + i \omega_{\text{I}_{\text{Initial}}}$. One can solve the QNMs by standard shooting method.
%{as
%%
%\begin{equation}
%\{B=0\}  \xrightarrow[\omega_{\text{Initial}}=\omega_{\text{R}_{\text{Initial}}} + i \omega_{\text{I}_{\text{Initial}}}]{\text{Shooting}} \{\omega_{\text{QNMs}}\}.
%\end{equation}
%%
%}

Numerically, the error includes three parts in general. One comes from the finite difference when we solve the differential equation numerically, which can be suppressed by using higher order methods or smaller step-size. The second one comes from the fact that we have to set two finite cut-off at the horizon and infinite boundary. The third part comes from the float-point error of computer. We will show in following that, if imaginary part of QNMs is positive, errors of the second and third parts lead to the computational complexity will increase exponentially if we improve the desired accuracy of QNMs.

Instead of considering the boundary condition~\eqref{wavebound}, we firstly consider a general boundary-valued eigenvalue problem with following boundary condition for the radial equation Eq.~\eqref{derad}
\begin{equation}\label{wavesol}
\tilde{R} \varpropto \left\{
\begin{aligned}
&\te^{-i \omega y } + A \te^{i \omega y} \quad \quad  &&y \to -\infty \\
&\te^{i \omega y} + B \te^{-i \omega y}  \quad \quad  &&y \to \infty \, ,
\end{aligned}
\right.
\end{equation}
where $A, B$ are two complex constants. For a pair of given specific $A$ and $B$, the radial equation Eq.~\eqref{derad} can numerically give a complex frequency $\omega=\omega_{\text{R}}+i \omega_{\text{I}}$. Therefore, we can interpret a series of complex frequency as an analytical function with respect to $A$ and $B$, denoted as $\omega(A, B)$. Theoretically, it is obvious that the QNMs arise from vanishing both $A$ and $B$, $\omega_{\text{QNM}}=\omega(0,0)$, matching the boundary condition Eq.~\eqref{wavebound}. However, in the shooting method in numerics, $A, B$ cannot exactly vanish due to floating-point error. Practically, in order to solve QNMs in numerics, the real boundary condition in the numerical computations is
\begin{equation}\label{wavesol2}
\tilde{R} \varpropto \left\{
\begin{aligned}
&\te^{-i \omega y } + \epsilon \quad \quad  &&y \to -\infty \\
&\te^{i \omega y} + \epsilon \quad \quad  &&y \to \infty \, ,
\end{aligned}
\right.
\end{equation}
where $\epsilon$ denote the floating-point error in numerics. For a given complex frequency, comparing with Eq.~\eqref{wavesol} and Eq.~\eqref{wavesol2}, one can find that it is equivalent to set a pair of nonzero $A$ and $B$ and
%
%\begin{equation}
%\epsilon \sim \max\left\{\te^{-\omega_{\text{I}}y_{-\infty}}, \te^{\omega_{\text{I}}y_{\infty}} \right\}
%\end{equation}
%
%
\begin{equation} \label{Aep}
A \sim \epsilon \te^{\omega_{\text{I}}y_{-}} \, ,\quad B \sim \epsilon \te^{-\omega_{\text{I}} y_{+}} \, ,
\end{equation}
where $y_{\pm }$ are numerical cutoff at both negative infinity and positive infinity respectively. Thus, for a computer which has fixed machine precision, the floating-point error restricts our ability to find the QNMs in arbitrary precision. To improve the accuracy of finding QNMs, we have to increase the machine precision of computer. In general, $\omega(A, B)$ can be expanded as following approximation upon the 1st order of $A$ and $B$
\begin{equation}
\omega(A,B)= \omega_{\text{QNM}} +\frac{\partial\omega}{\partial A} A + \frac{\partial \omega}{\partial B} B + \cdots\,.
\end{equation}
We thus can obtain following estimation on the error of finding QNMs
\begin{equation}
\omega_{\text{err}}:=|\omega(A,B)-\omega_{\text{QNM}}|\sim\max\left\{\left|\frac{\partial \omega}{\partial A} \right|\left|A\right|,\left|\frac{\partial \omega}{\partial B}\right| \left| B\right| \right\}.
\end{equation}
denotes the numerical error associated with the QNMs. As it is reasonable to assume that $\omega(A,B)$ is the analytical function of $A$ and $B$, so $\partial\omega/\partial A$ and $\partial\omega/\partial B$ are both finite. Thus,it can be concluded that the error is controlled by $A$ and $B$, i.e. $\epsilon \te^{\omega_{\text{I}}y_{-}}$ and $\epsilon \te^{-\omega_{\text{I}} y_{+}}$.

For a series of  unstable mode, $\omega_{\text{I}} > 0$, triggering on exponential growth as time evolution, from Eq.\eqref{Aep} one can find that an effective working precision with given $\epsilon$ both $A, B$ will exponentially decay as large cutoff, ensuing numerical accuracy that $\omega_{\text{err}}$ is convergence in finite computation time. Therefore, the error caused by float-point is suppressed exponentially and we only need to care about the errors caused by finite difference and cut-off. In this case, the shooting method can catch the unstable mode efficiently.

However, it does not work for stable modes $\omega_{\text{I}} < 0$, in which $\psi_2$ will exponential damp as time evolution. In this case, we will suffer a contradiction that if we suppose $\omega_{\text{err}}$ is convergence to guarantee accuracy, the required working precision, related to computing time, will grow exponentially as the growth of cutoffs $y_{\pm}$, while in order to obtain a precision-guarantee QNMs, we need to set $y_{\pm}$ as large as possible to reduce cutoff error. Said another specific way, if one improve ten times of the cutoff, $y_{\pm} \to 10 y_{\pm}$, the required working precision has to be improved up to $\te^{10 \omega_{\text {I}}}$ to ensure computation precision. Consequently, the numerical approach cannot catch the stable modes efficiently.

To illustrate our analysis more clearly, we take the negative P\"{o}schl-Teller potential as an example in Appendix.~\ref{PT}, in which shows how the shooting method can catch unstable modes efficiently.  Based on our illustration above, we adopt the numerical approach to calculate the unstable mode $\omega_{\text{I}} > 0$ while the WKB approximation method to calculate the stable mode $\omega_{\text{I}} < 0$.

\begin{figure}[hbpt]
  \centering
  % Requires \usepackage{graphicx}
  \includegraphics[width=0.5\textwidth]{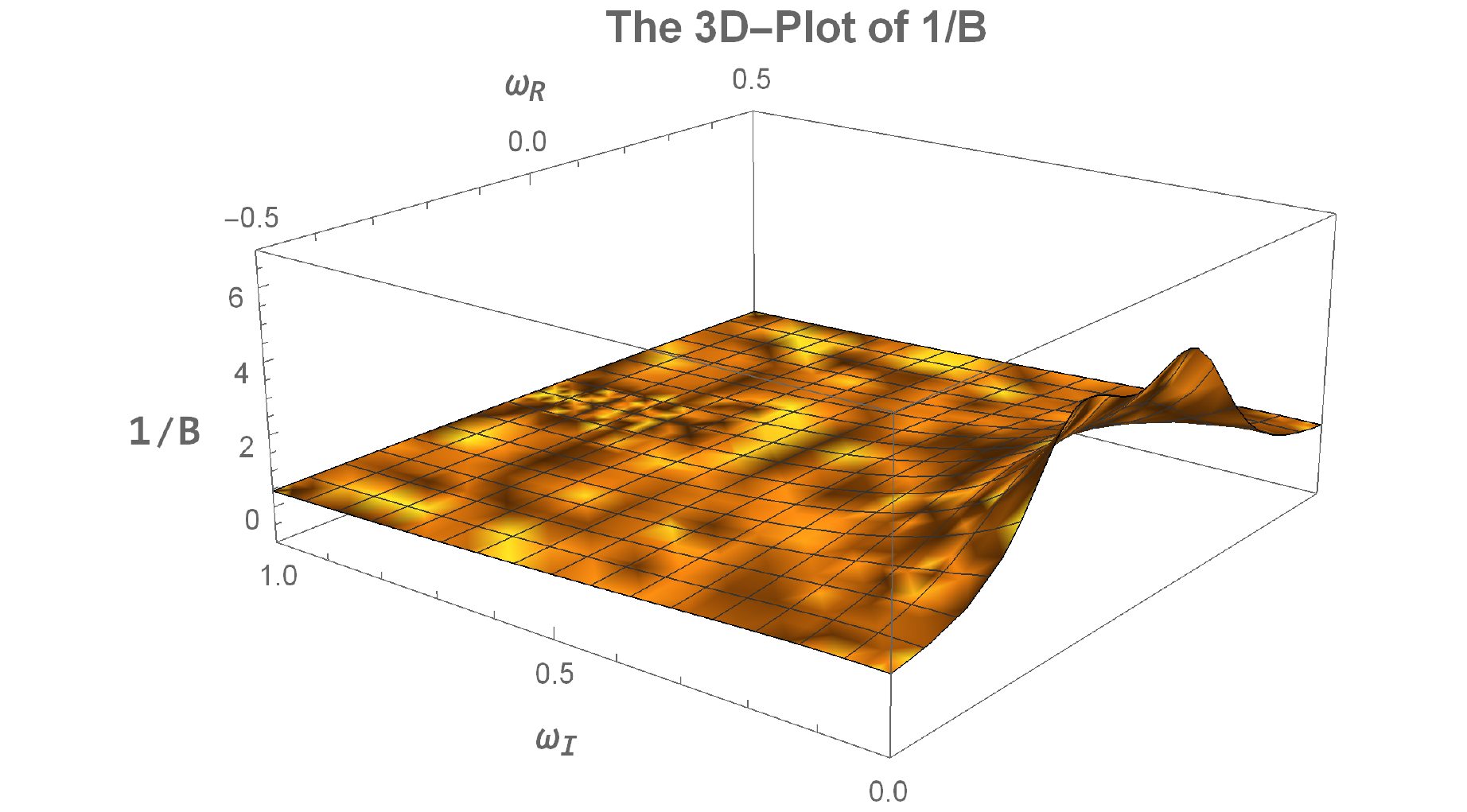}
  \caption{3D-plot associated with $\frac{1}{B}$ with $(\omega_{\rm R}, \omega_{\rm I})$ plane , taking the numerical solution $\{c=0.1, \psi_0 \approx 0.1988\}$ with $\{r_h=1, \phi_1=0.8, m=\frac{1}{100}, q=\frac{11}{10}m\}$ as an example.}\label{numpic1}
\end{figure}

To end this section, we adopt the numerical solution, denoted as $\{c=0.1, \psi_0 \approx 0.1988\}$ with $\{r_h=1, \phi_1=0.8, m=\frac{1}{100}, q=\frac{11}{10}m\}$, to investigate whether there exist unstable mode within QNMs. Following the above procedure, we show a 3D-plot associated with $1/B$ with $(\omega_{\rm R}, \omega_{\rm I})$ plane in Fig.~\ref{numpic1}. One can easily observe that there does not exist any peak of singularity which indicates that the existence of unstable mode within QNMs. Furthermore, we adopt some another numerical solutions $\{c_{i}, \psi_{0_{i}}\}$ to investigate whether there exist unstable mode and cannot observe any other unstable mode within QNMs. Before we give a conclusion that the scalarized charged black hole is stable against a neutral scalar field at linear level, in the following section we also adopt the WKB approximation method to calculate the stable mode in QNMs.

\subsection{The WKB Approximation Method}
At first, we give brief introduction on calculating the QNMs by the WKB approximation method. The WKB approximation method was firstly used to calculate the QNMs of Schwardzchild black hole by Schutz and Will \cite{Shusz}. It then was developed to the 3rd order by Iyer and Will \cite{Iyer:1986np,Iyer:1986nq} and further 6th order by R.A. Konoplya \cite{Konoplya:2003ii}. Recently it has been extended to the 12nd order by Matyjasek and Opala \cite{Matyjasek:2017psv}. In this paper, we adopt the the 3rd order WKB approximation method to analyze the QNMs of the scalarized charged black hole. We point out again the we have used numerical method in the last subsection verified that there is no unstable QNMs.

\begin{figure}[h]
  \centering
  % Requires \usepackage{graphicx}
  \includegraphics[width=0.3\textwidth]{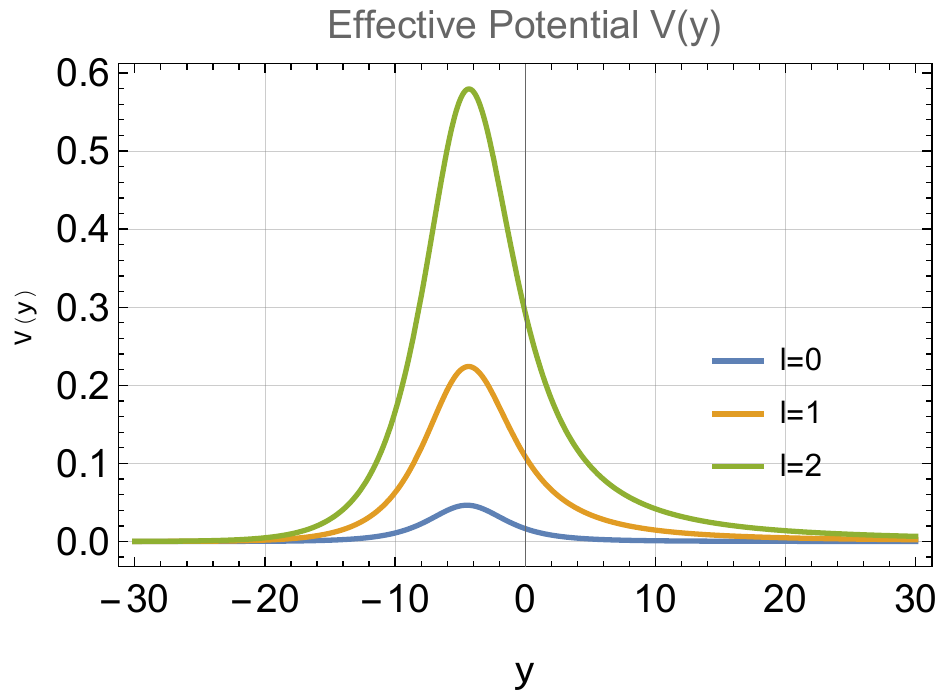}
  \caption{ The effective potential under the tortoise coordinate $V(y)$. }\label{effV1}
\end{figure}

Recall the radial function Eq.~\eqref{derad}, the effective potential $V(y)$ gives
\begin{equation}\label{effVy}
V(y)=\frac{l(l+1)}{r(y)^2}f(y) \te^{-\chi(y)}+\frac{1}{2r(y)}\frac{d}{dr}\left(f(y)^2 \te^{-\chi(y)}\right)
\end{equation}
We firstly consider the numerical scalarized charged black hole solution $\{c=0.1 ,\psi_0=0.1987\}$ as an example, found in Sec.~\ref{thermo}. Substituting the numerical solution in Eq.~\eqref{effVy}, we present a plot of $V(y)$ taking $l=0,1,2$ as example in Fig.~\ref{effV1}. Furthermore, one can numerically read off the local maximal value $V_0(y_0)$ from Eq.~\eqref{effVy} (See also Table.~\ref{V0y0})
\begin{table}[hbtp]
\centering
\begin{tabular}{c|c|c|c}
  \hline
  \hline
  & $l=0$ & $l=1$ & $l=2$ \\
  \hline
 $V_0$ & $ 0.04641$ & 0.2240 & $0.5794$  \\
  \hline
  $y_0$& $ -4.485$ & $ -4.355$ & $-4.327$         \\
 \hline
 \hline
\end{tabular}.
\caption{The local maximal $V_0$ of the effective potential $V(y)$ under the numerical scalarized charged black hole solution $\{c=0.1, \psi_0=0.1978 \}$. } \label{V0y0}
\end{table}
We then proceed to compute the QNMs using the WKB approximation method up to the 3rd order. To maintain accuracy, we respectively take $l=0$ up to $n=2$ as well as $l=1,2$ up to $n=5$ as example. Recall the formula of the 3rd order WKB Approximation method, the QNMs have been given explicitly in Ref.~\cite{Iyer:1986np,Iyer:1986nq} as follow,
\begin{equation}
\omega^2=(V_0+\sqrt{-2V''_{0}} \Lambda_2)-i(n+\frac{1}{2})\frac{1}{\sqrt{-2 V_0''}}(1+\Lambda_3),
\end{equation}
where
\begin{eqnarray}\label{WKB}
\Lambda_2 &=& \frac{1}{\sqrt{-2 V_0''}}\left(\frac{1}{8}\left(\frac{V^{(4)}_0}{V_0''}\right)\left(\alpha^2+\frac{1}{4}\right)\right. \cr
&&\left.-\frac{1}{288}\left(\frac{V^{(3)}_0}{V_0''}^2(60\alpha^2+7) \right)\right), \nonumber  \\
~ \cr
\Lambda_3 &=& \frac{1}{-2V_0''}\left(\frac{5}{6912}\left(\frac{V^{(3)}_0}{V_0''}\right)^4(188\alpha^2+77)\right. \cr
&&-\frac{1}{384}\left(\frac{V^{(3)2}_0V^{(4)}_0}{V''^3_0}\right) (100\alpha^2+51)\cr
&&+\frac{1}{2304}\left(\frac{V^{(4)_0}}{V''_0}\right)^2(68\alpha^2+67) \cr
&&+\frac{1}{288}\left(\frac{V'''_0 V^{(5)}_0}{V''^2_0} \right) (28\alpha^2+19) \cr
&&- \left.\frac{1}{288}\left(\frac{V^{(6)}_0}{V''_0}\right)(4\alpha^2+5) \right)\, ,
\end{eqnarray}
where $V_0^{(n)}$ denotes the $n$-th derivative of $V(y)$ with respect to $y$ located on $y_0$ and $\alpha=(n+\frac{1}{2})$. In Eq.~\eqref{WKB}, $\Lambda_2$ and $\Lambda_3$ denote the second order and the third order approximation associated with the WKB method, respectively Plugging the specific numerics into Eq.~\eqref{WKB}, we then show the results of $n \leq l+1$ in which $l=0, 1,2 $ respectively for maintaining precision (See also the Table.~\ref{QNMs}). Since the WKB method is sufficient in high-lying mode but may not be sufficient enough in low-lying mode, we do not consider the high overtone case. In Table.~\ref{QNMs}, we show the numerical results of QNMs up to 3rd and the associated relative errors $\Delta$, defined as $\Delta = |\omega-\omega_2|/|\omega|$ where $\omega_2$ denotes the numerical QNMs given by the second order WKB method. One can observe that the imaginary part of the complex frequency $\omega_{\rm I}$ is always negative, indicating that the numerical solution $\{c_0, \psi_{0_0}\}$ is stable against the scalar perturbation.
\begin{table}[hbtp]
\centering
\begin{tabular}{c|c|c|c}
  \hline
  \hline
  $n$ & $l=0$ & $l=1$ & $l=2$ \\
  \hline
 $0$ & \tabincell{c}{$0.146-0.125 i$ \\ $~24.8\%$ } & \tabincell{c}{$0.450-0.112 i$ \\$  2.03\%$ } &  \tabincell{c}{$0.748-0.112 i$  \\ $0.488 \% $ } \\
  \hline
  $1$& \tabincell{c}{ $0.104-0.409i $  \\ $~22\% $ } & \tabincell{c}{$ 0.419 - 0.350 i  $ \\$ ~6.24\% $ }  & \tabincell{c}{ $0.728- 0.340 i$   \\ $1.90 \%$} \\
  \hline
 $2$&  &  \tabincell{c}{$0.371 - 0.601 i $ \\ $~10\% $ } & \tabincell{c}{ $0.692-0.575 i$ \\  $ 4.23 \%$ } \\
   \hline
 $3$& &  &  \tabincell{c}{$0.646-0.816 i$ \\ 7.07 \% } \\
 \hline
 \hline
\end{tabular}
\caption{The QNMs of the scalarized charged black hole with $\{ c=0.1, \psi_0=0.1988\}$. In each cell we show the numerical results of QNMs and associated relative error $\Delta = |\omega-\omega_2|/|\omega|$. }\label{QNMs}
\end{table}

Recall a series numerical scalarized charged black hole solution, denoted by $\{c_{i}, \psi_{0_i}\}$ in Sec.~\ref{micensem}, for without lost generality, we calculate $\omega_{\rm I}$ associated with $c_i$ by the WKB approximation method as well. We find that $\omega_{\rm I}$ are always negative, and do not change obviously as the growth of $c_i$, implying that the QNMs of scalarized charged black hole is less affected by the amplitude of the non-linear potential. Furthermore, we also have checked other different parameters and found the similar results. This suggests that the scalarized charged black hole should be kinetically stable under a neutral perturbation.\footnote{Strictly speaking, the QNMs do not form a complete bases in mathematics, so above analysis does not cover all possible perturbations in mathematics. }

\section{Conclusion}\label{Conclu}
In this paper, we consider the Einstein-Maxwell theory minimally coupled with a non-linear complex field. Considering an appropriate boundary condition for scalarized black hole in asymptotic flat spacetime. We at first briefly list our main result:
\begin{itemize}
\item{For general non-linear semi-definite potential, we prove that the scalarization cannot result from a continuous phase transition. }

\item{Treating the scalarized black hole as a thermodynamical system, we observe that the discontinuous scalarization on RN black hole will not happen in grand canonical ensemble but will happen in both microcanonical ensemble and canonical ensemble.}

\item{We also find that neutral scalar perturbation will not trigger kinetic instability associated with the scalarized charged black hole by means of analysing the QNMs using numerical method and the WKB approximation method.}
\item{As a by-product, we use numerical results to give negative answer to Penrose-Gibbons conjecture and suggest two new versions of Penrose inequality in charged case.}
\end{itemize}

In detail, motivated by the numerical solution as a counterexample for the no-hair theorem given in \cite{Hong:2020miv}, we investigate the thermodynamic stability of the scalarized charged black hole, compared with the RN black hole in various ensembles. It needs to note that it is more suitable to choose grand canonical ensemble for a black hole in astrophysics. However, in this paper, we still take microcanonical ensemble and canonical ensemble into account as theoretical research interests.

In microcanonical ensemble, ADM mass $M$ and total charge $Q$ are fixed and the phase transition will happen towards the direction which increases the entropy, namely the radius of the event horizon. Giving specific $M$ and $Q$ and working with $(Q/M,\Delta r_h/M)$ plane, we find that it is possible the scalarized charged black hole is more stable than the RN black hole in thermodynamics. Particularly, our numerical results imply that the scalarized black hole is more stable than RN black hole when temperature is low enough and the near extremal RN black hole will always transit into scalarized black hole via a first order phase transition.

As to in the canonical ensemble and grand canonical ensemble, the stability in thermodynamics requires the minimal of the Helmholtz free energy $F(T, Q)$ with fixing Hawking temperature $T$ and total charge $Q$, and the Gibbs free energy $G(T,\mu)$ with fixing $T$ and chemical potential $\mu$ respectively. Taking probe limit, we firstly present a proof that the RN black hole is more thermodynamically stable in both canonical ensemble and grand canonical ensemble. Following the similar procedure in microcanonical ensemble, giving specific $T$ and $Q$ and working with $(T/Q, \Delta F/Q)$ and $(Q/T, \Delta F / T)$ plane, we also find it is possible that the scalarized charged black hole is more stable in thermodynamics in canonical ensemble and so the RN black hole may  spontaneously scalarize via a first order phase transition in canonical ensemble. However, in grand canonical ensemble, we find that RN black hole always have smaller Gibbs free energy and so is more stable than scalarized black hole, which implies that the RN black hole will not spontaneously scalarize in grand canonical ensemble.

Finally, we study the kinetic stability of the scalarized charged black hole against scalar perturbation. Due to the static and spherical spacetime background,  we firstly reduce the master equation of the perturbative scalar to $1$ dimension in frequency domain. Given pure ingoing wave condition at the horizon as well as pure outgoing wave condition at the spatial infinity, a series of complex frequency $\omega=\omega_{\rm R}+ i \omega_{\rm I}$ is picked out, namely the quasi-normal modes (QNMs) of  black hole. Through numerical error analysis, we claim that the shooting method in numerics can efficiently catch the unstable mode $\omega_{\rm I}>0$, rather than the stable mode $\omega_{\rm I}<0$. Therefore, we calculate the unstable mode within QNMs by the shooting method in numerics and conclude that there does not exist any unstable mode within the QNMs associated with the scalarized charged black hole. In addition, since the WKB approximation method can effectively catch the stable modes, rather than the unstable mode, within the QNMs, we adopt the 3rd order WKB approximation method to compute the QNMs of the scalarized charged black hole and find that the imaginary part of the QNMs $\omega_{\rm I}$ is always negative. Therefore, we concluded that the scalarized charged black hole should be kinetically stable under a neutral perturbation. As further discussion on future work, it is worth to investigate whether other perturbation, for instance, vector perturbation or tensor perturbation will trigger on kinetic instability. We thus will keep focusing on this topic in our future work.

As a by-product of our numerical construction of scalarized black hole, we also definitely give negative answer to a long-standing conjecture named Penrose-Gibbons conjecture. Particularly, we find that the the total charge can be larger than the ADM mass but the temperature is still positive in scalarized black hole. Based on our numerical results, we propose two new generalizations of Penrose inequality in charged case and numerically verify their correctness in our model. Moreover, we will give more discussion in our future works.

\acknowledgments
We are grateful to Hong L\"{u}, De-Cheng Zou and A. Zhidenko for useful discussions and Wen-Di Tan, Shi-Fa Guo and Ze Li for proofreading.

\appendix
\section{Efficiency of Numerical Method for Searching Unstable Modes: Negative P\"{o}schl-Teller Potential }\label{PT}
\begin{figure}[h]
  \centering
  % Requires \usepackage{graphicx}
  \includegraphics[width=0.3\textwidth]{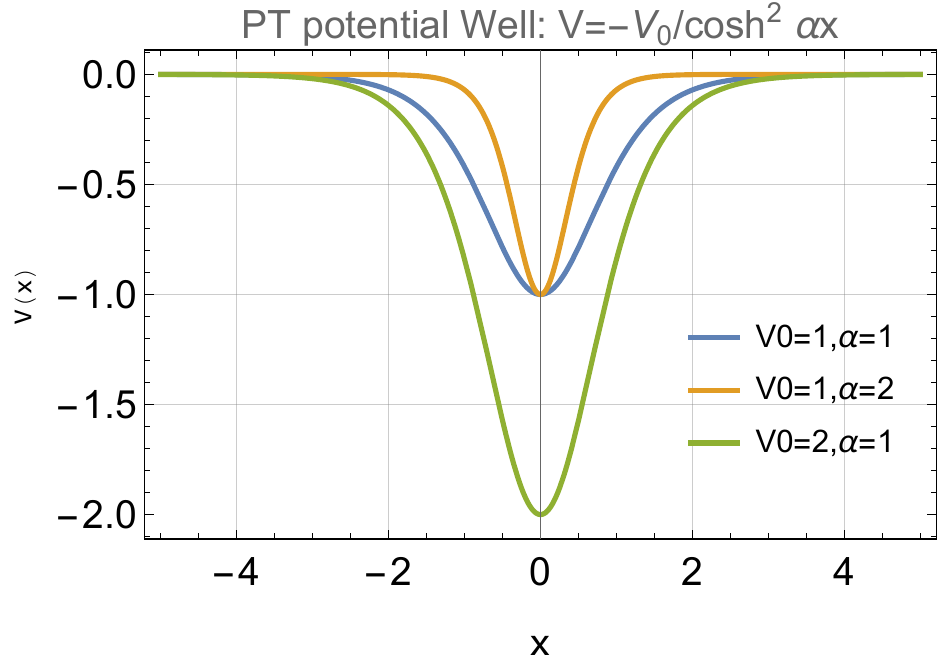}
  \caption{The P\"{o}schl-Teller potential well $V(x)=V_0/\cosh^2 \alpha x$.}\label{PTp}
\end{figure}

\begin{figure}[hbtp]
\centering
\begin{minipage}{0.48\textwidth}
\centerline{\includegraphics[width=0.6\textwidth]{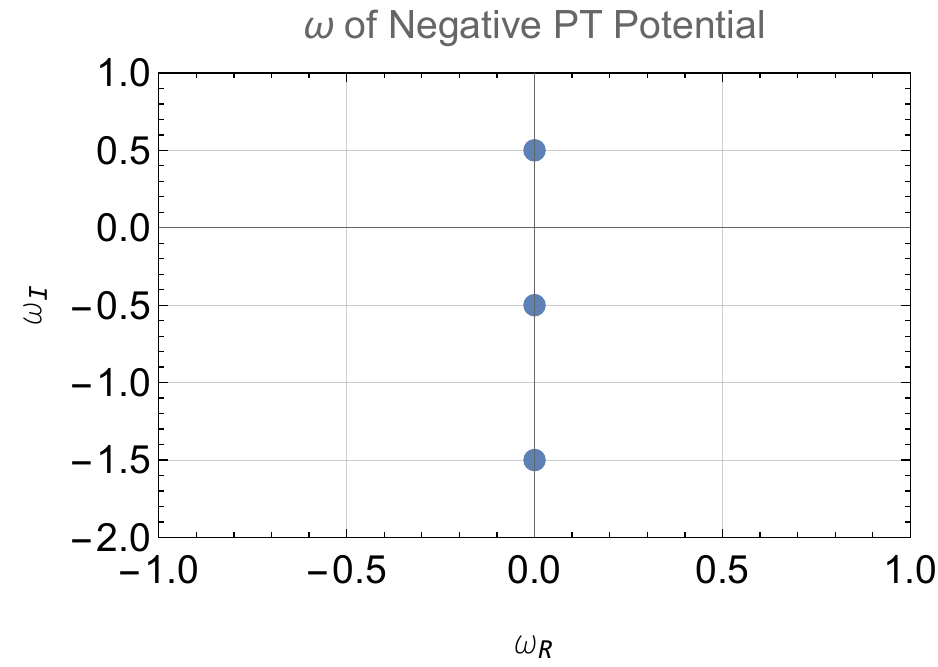}}
\end{minipage}
\hfill
\begin{minipage}{0.48\textwidth}
\begin{tabular}{r r}
\hline
  \hline
  $n$ & $\omega_{\text{I}}$ \\
  \hline
 $0$ & $1/2$  \\

 $1$& $-1/2$      \\

 $2$& $-3/2$ \\
 \hline
 \hline
\end{tabular}
\end{minipage}
\caption{The "QNMs" of negative PT potential }\label{PTome}
\end{figure}

In this section, we shall illustrate that the numerical method is efficient for searching unstable QNMs Negative P\"{o}schl-Teller Potential as example. Following the procedure given in Ref.~\cite{Ferrari:1984zz}, the "QNMs"(the bound frequency) of the negative PT potential can be analytically solved and we then turn to verify it by the shooting method in numerics in this section. At first, we consider the equation of motion associated with a one-dimension wave under a P\"{o}schl-Teller potential well (PT potential) in frequency domain can be written as
\begin{equation}\label{eqPT}
\frac{\td^2 \Psi}{\td x^2}+(\omega^2-V_{\text{PT}}(x))\psi(x)=0\, ,
\end{equation}
where
\begin{equation}\label{VPT}
V_{\text{PT}}=\frac{V_0}{\cosh^2 \alpha x}.
\end{equation}
In general, the case with $V_0 >0$, in which the unstable mode $\Im(\omega)>0$ does not exist, has been under consideration in \cite{Ferrari:1984zz}. To stimulate the unstable mode, in this paper we consider the so-called negative PT potential case where $V_0<0$, $V_0=V(0)$ and $\alpha=\left.-\frac{1}{2V_0}\left(\frac{\td^2 U}{\td x^2} \right)\right|_{x=0}$ with shape as potential well(See also Fig.~\ref{PTp}).
The "QNMs" then are given as
\begin{equation}
\omega=-i \alpha \left(n+\frac{1}{2} \right) \pm i \alpha \sqrt{\frac{1}{4}-\frac{V_0}{\alpha^2}} \, , \quad V_0 <0 \, ,
\end{equation}
where $n=0, 1, 2, \cdots$. One can find that all $\omega$ lie on the imaginary axis, where $\omega_{\text{I}}<0$ stands for a stable mode and $\omega_{\text{I}}>0$ denotes an unstable mode. For convenience, we specifically taking parameters that $\alpha=1, V_0=-\frac{3}{4}$, a series of frequency is shown in Fig.~\ref{PTome}. It is obvious to find that when $n\geq 0$, there only exist a single unstable mode lying on $\omega=0+i\frac{1}{2}$. We show our result in Fig.~\ref{PTome}.

\begin{figure}[hbpt]
  \centering
  % Requires \usepackage{graphicx}
  \includegraphics[width=0.35\textwidth]{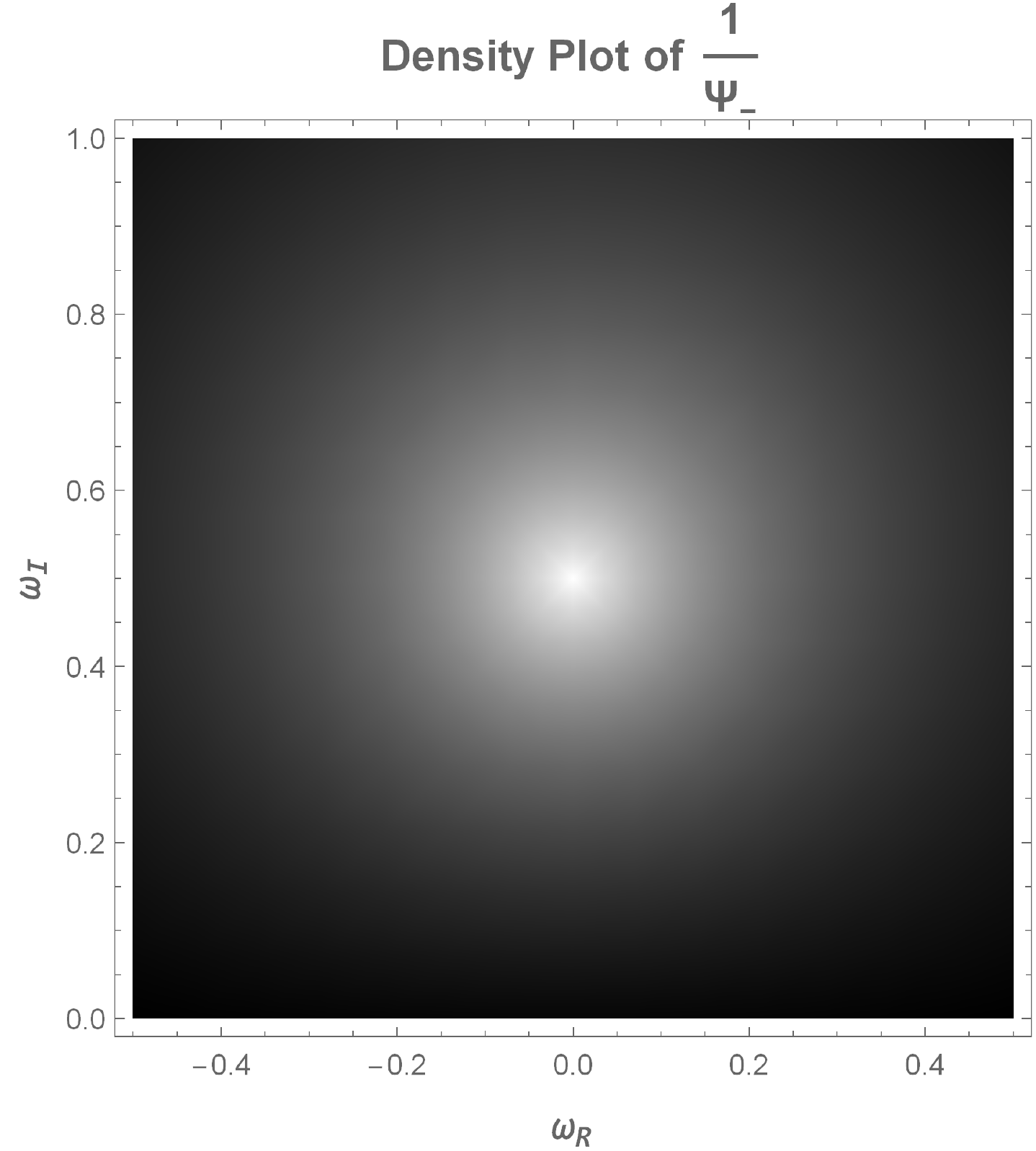}
  \caption{The density plot associated with $1/\Psi_-$ of negative PT potential in numerics.}\label{PTpic}
\end{figure}

In following, we verify the consistency of our illustration in Sec.~\ref{numerr} by the shooting method. In the case of the negative PT potential, the superfluousness of tortoise coordinate indicates the absence of the differential equation associated with the definition of tortoise coordinate. Therefore, we establish a boundary valued problem including a second-order differential equation Eq.~\eqref{eqPT} and boundary condition related to QNMs
\begin{equation}\label{boundpt}
\Psi \varpropto \left\{
\begin{aligned}
&\te^{-i \omega x } \quad \quad  &&x \to -\infty \\
&\te^{i \omega x}   \quad \quad  &&x \to \infty \, .
\end{aligned}
\right.
\end{equation}
In practise, one can numerically integrate Eq.~\eqref{eqPT} with one-side boundary condition in $x \to -\infty$, then in $x \to \infty$ region the numerical integration will generally give
\begin{equation}
\Psi \varpropto \Psi_{+} \te^{i \omega x} + \Psi_{-} \te^{- i \omega x} \, , \quad x \to \infty \, .
\end{equation}
Furthermore, we have
\begin{equation}
\Psi_{-}=\frac{\te^{i \omega x}}{2 \omega}\left(\omega \Psi + i \Psi'  \right)\, , \quad x \to \infty \, .
\end{equation}
Comparing the boundary condition associated with the QNMs and interpreting $\Psi_-$ as an analytical function with respect to $\omega$, one can easily find that vanishing $\Psi_{-}$ will give QNMs, namely
\begin{equation}
\Psi_{-}(\omega_{\rm QNMs})=0 \, .
\end{equation}
To calculate the QNMs of negative PT potential, we firstly draw a density plot of $\frac{1}{\Psi_{-}}$ in $(\omega_{\rm R}, \omega_{\rm I})$. As our illustration in Sec.~\ref{numerr}, the location of the QNMs  will become a bright spot as a singularity of our density plot related to $1/\Psi_{-}$ since $\omega_{\rm QNMs}$ indicate $\Psi_{-}=0$. From the density plot one can read an approximate value of $\omega$ as an initial value, namely $\omega=\omega_{\rm R_{\rm Initial}}+ i \omega_{\rm I_{\rm Initial}}$, then the QNMs can be solved by standard shooting method
\begin{equation}
\{\Psi_-=0\}  \xrightarrow[\omega_{\text{Initial}}=\omega_{\text{R}_{\text{Initial}}} + i \omega_{\text{I}_{\text{Initial}}}]{\text{Shooting}} \{\omega_{\text{QNMs}}\}.
\end{equation}
Choosing $V_0=\frac{3}{4}$ and $\alpha = 1$. We show a density plot associated to $1/\Psi_-$ in Fig.~\ref{PTpic}, working in $(\omega_{\text{R}}, \omega_{\rm I})$ with $\omega_{\rm R} \in [-0.5, 0.5]$ as well as $\omega_{\rm I} \in [0, 1]$.

One can easily observe that there exist a bright spot near $\omega_{\rm R}=0$ and $\omega_{\rm I} \approx 0.5$. Using shooting method, we obtain a numerical result of QNMs that
\begin{equation}
\omega_{\rm QNMs}\approx 5.189 \times 10^{-10} + i(0.5 + 5 \times 10^{-5}) \, ,
\end{equation}
indicating that within the allowable region in numerical error, the numerical result of the unstable mode is consistent with the value given by theoretical calculation. Therefore, it is concluded that the shooting method in numerics can efficiently catch the unstable mode within the QNMs.

\section{Shooting Method for Thermodynamical Stability of Scalarized charged black hole}\label{numapp}
In this section, we shall present our explicit numerical results of shooting method for studying the thermodynamical stability associated with the scalarized charged black hole.
\subsection{Microcanonical Ensemble}\label{micapp}
By fixing the independent parameters as $\{r_h=1, \phi_1=0.8, m=0.01, q=\frac{11}{10}m , r_{\infty}=1000 \}$, we set the $\{ {c_0}, {\psi_{0_{0}}}\}$, shown in Fig.~\ref{fig1}, as seed solution, then find next solution $\{{c_{\pm 1}},\psi_{0_{\pm 1}}\}$ taking $\{ {c_0}=0.1, {\psi_{0_{0}}} \approx 0.1988\}$ as the initial value in the shooting method. For maintaining numerical stability we set the step size $c_{ \pm i}=c_{\pm(i+1)} \pm \frac{1}{1000}$. Starting with $\{c_0=0.1, \psi_0 \approx 0.1988  \}$, we then obtain a class of numerical scalarized charged black hole represented by $\{ c_{\pm i} , \psi_{0_{ \pm i}}\}$ where $i=1,2,3 , \cdots$. Given specific $M_i$ and $Q_i$, the map Eq.~\eqref{map1} indicates one can leave three parameters free and fix two under the constrains $\{M=M_i, Q=Q_i,\psi_+=0 \}$ (The last one is from the constrain Eq.~\eqref{constraint1} for spontaneous scalarization). Said another way, we can obtain a class of numerical scalarized charged black hole solution with specific $M_i$ and $Q_i$ by the shooting method. Concretely, fixing $\{c \approx 0.01220, q=1.1m \}$, we further obtain a series of numerical charged black hole solutions with scalar hair, denoted by three parameters $\{r_{h_{i}}, \phi_{1_{i}}, \psi_{0_{i}} \}$, for given $\{M=M_i=0.9676, Q=Q_i,\psi_+=0 \}$. Following the trick we set up above, the parameters of previous numerical solution $\{r_{h_{i-1}}, \phi_{1_{i-1}}, \psi_{0_{i-1}}\}$  will be used as the initial value for shooting the next solution $\{r_{h_{i}}, \phi_{1_{i}}, \psi_{0_{i}}   \}$ under $\{M=M_i = 0.9676, Q=Q_i ,\psi_+=0 \}$. It can be written as
\begin{eqnarray}
&&\{M=M_i = 0.9676, Q=Q_i ,\psi_+=0 \}   \xrightarrow[\{r_{h_{i-1}}, \phi_{1_{i-1}}, \psi_{0_{i-1}} \}]{\text{shooting}} \cr
&&\{r_{h_{i}}, \phi_{1_{i}}, \psi_{0_{i}} \}\,.
\end{eqnarray}
Starting with the seed solution $\{ r_{h_{0}}=1, \phi_{1_{0}}=0.8, \psi_{0_{0}} \approx 0.02473 \}$ together with fixing parameters $\{c =0.01220, \psi_0=0.02473 \}$, we eventually obtain a series of numerical scalarized charged black hole solution with $M=0.9676, Q=Q_i$, where we set the step size $Q_{i+1}=Q_i - \frac{1}{1000}, Q_0=0.9676$ for maintaining accuracy.

\subsection{Canonical Ensemble}\label{canapp}
To search a good seed solution, we firstly work with $(c, \Delta F )$ plane. From the plot presented in the first panel of Fig.~\ref{fig3}, we find that there does exist a zero point of $\Delta F$ in interval $[0.020, 0.023]$. Adopting the interpolation as well as the shooting method, we search another numerical seed solution $\{c \approx 0.02148, \psi_0 \approx0.04350 \}$ with $T=0.02858$ and $Q=1.318$ , sharing the same $F(T,Q)$, $T$ and $Q$ with the corresponding RN black hole.With the seed solution and the map Eq.~\eqref{map1}, for given specific $T_i$ and $Q_i$, we can obtain a series of numerical scalarized charged black hole solution under the constrains $\{T=T_i, Q=Q_i, \psi_+=0\}$ by the shooting method. Working with $(T, \Delta F)$ plane and $(Q, \Delta F)$ plane respectively and choosing the fixed parameters $\{c=0.02148, q=1.1m, m=0.01\}$, we start with the seed solution $\{r_{h_0}=1, \phi_{1_0}=0.8, \psi_{0_0} \approx 0.04350 \}$ and respectively obtain two series of numerical solutions $\{r_{h_i}, \phi_{1_i}, \psi_{0_i}\}$ with $T=T_i=0.02858, Q=Q_i$ and $Q=Q_i=1.318,T=T_i$ by shooting method,
\begin{eqnarray}
&& \{T=T_i = 0.02858, Q=Q_i ,\psi_+=0 \}   \xrightarrow[\{r_{h_{i-1}}, \phi_{1_{i-1}}, \psi_{0_{i-1}} \}]{\text{shooting}} \cr
&&\{r_{h_{i}}, \phi_{1_{i}}, \psi_{0_{i}} \}\, ,  \\
~\cr
&& \{T=T_i, Q=Q_i=1.318 ,\psi_+=0 \}   \xrightarrow[\{r_{h_{i-1}}, \phi_{1_{i-1}}, \psi_{0_{i-1}} \}]{\text{shooting}}  \cr
&& \{r_{h_{i}}, \phi_{1_{i}}, \psi_{0_{i}} \}\, .
\end{eqnarray}
For holding accuracy, we set the step size that $Q_{i}=Q_{i-1}-\frac{1}{100}$ from $Q_0=1.318$ and $T_{\pm i}=T_{\pm(i-1)}\pm \frac{1}{1000}$ from $ T_0=0.02858$ respectively. In the case of canonical ensemble, one can read off $F(T_i,Q_i)$ of the corresponding RN black hole with temperature $T_i$ and total charge $Q_i$ from Eq.~\eqref{Frn} and Eq.~\eqref{rnTQ}. Therefore, we numerically build a relation between $\Delta F (T_i, Q_i) = F(T_i, Q_i)- F_{\rm RN}(T_i, Q_i)$ and $T$ or $Q$.
~\\

\subsection{Grand Canonical Ensemble}\label{graapp}
We firstly consider the smaller $c=0.00101$ case. Recall the approach we adopted previously, fixing parameter $\{c=0.00101, q=1.1m\}$ and starting with the seed solution $\{r_{h_0}=1, \phi_{1_0}=0.8, \psi_{0_0} \approx 0.002047\}$ with $\{T_0=0.02864, \mu=0.8000\}$, we obtain two series of numerical solution $\{r_{h_i}, \phi_{1_i}, \psi_{0_i}\}$ under the constrains $\{T=T_i, \mu=\mu_i, \psi_+=0 \}$ by shooting method,
\begin{eqnarray}
&& \{T=T_i = 0.02864, \mu=\mu_i ,\psi_+=0 \}   \xrightarrow[\{r_{h_{i-1}}, \phi_{1_{i-1}}, \psi_{0_{i-1}} \}]{\text{shooting}} \cr
&& \{r_{h_{i}}, \phi_{1_{i}}, \psi_{0_{i}} \}\, ,  \\
~\cr
&& \{T=T_i, \mu=\mu_i=0.8000 ,\psi_+=0 \}   \xrightarrow[\{r_{h_{i-1}}, \phi_{1_{i-1}}, \psi_{0_{i-1}} \}]{\text{shooting}}  \cr
&& \{r_{h_{i}}, \phi_{1_{i}}, \psi_{0_{i}} \}\, .
\end{eqnarray}
Here we set the step size that $T_i=T_{i-1}+\frac{1}{100}$, $T_{-i}=T_{-(i-1)}-\frac{1}{100}$ while $\mu_{\pm i}=\mu_{\pm (i-1)} \pm \frac{1}{100}$ for maintaining numerical precision and stability.

As to the case of $c=0.1$, we briefly illustrate our results. Following the procedure we used above and starting with the seed solution $\{r_{h_0}=1, \phi_{1_0}=0.8, \psi_{0_0}=0.1987 \}$ with temperature $T_0=0.02721$ and chemical potential $\mu_0=0.8114$, we obtain another two series of numerical solutions $\{r_{h_i}, \phi_{1_i}, \psi_{0_i}\}$ under the constrains $\{T=T_i, \mu=\mu_i, \psi_+=0 \}$, where the step size is set as $\mu_{\pm i}=\mu_{\pm(i-1)} \pm \frac{1}{100}$ and $T_{i}=T_{i-1}+\frac{1}{100}$, $T_{-i}=T_{-i-1}-\frac{1}{500}$. For every $T_i$ and $\mu_i$, we pick out $\Delta G = G(T,Q)-G_{\rm RN}(T,Q)$ from Eq.~\eqref{Frn} and Eq.~\eqref{rnTQ}.

\end{document}